\documentclass[traditabstract,printer]{aa}

\usepackage{graphicx}
\usepackage{txfonts}
\usepackage{natbib}
\bibpunct{(}{)}{;}{a}{}{,}

\begin{document}

\title{Recurring millimeter flares as evidence for star-star magnetic reconnection events in the DQ Tau PMS binary system}

\author{
 D.~M. Salter \inst{1}
 \and
 \'{A}. K\'{o}sp\'{a}l \inst{1}
 \and
 K.~V. Getman \inst{2}
 \and 
 M.~R. Hogerheijde \inst{1}
 \and 
 T.~A. van Kempen \inst{3}
 \and
 J.~M. Carpenter \inst{4}
 \and
 \\G.~A. Blake \inst{5}
 \and
 D. Wilner \inst{3}
}

\institute{Leiden Observatory, Leiden University, PO Box 9513, 2300 RA Leiden, The Netherlands \\ 
\email{demerese@strw.leidenuniv.nl} 
\and
Department of Astronomy \& Astrophysics, 525 Davey Laboratory, Pennsylvania State University, University Park, PA 16802, USA
\and
Harvard-Smithsonian Center for Astrophysics, 60 Garden Street, MS 78, Cambridge, MA 02138, USA   
\and
Department of Astronomy, California Institute of Technology, Mail Stop 105-24, Pasadena, CA 91125, USA
\and
Division of Geological and Planetary Sciences, California Institute of Technology, Mail Stop 150-21, Pasadena, CA 91125, USA
}

\date{Submitted 10 June 2010; accepted 02 August 2010}

\authorrunning{Salter et al.}

\titlerunning{Recurring millimeter flares as evidence for star-star magnetic 
reconnection events toward DQ Tau} 

 
\abstract{Observations of the T~Tauri spectroscopic binary \object{DQ Tau} in April 2008 captured an unusual flare at 3\,mm, which peaked at an observed maximum flux of $\sim$0.5\,Jy (about 27 times the quiescent value). Here we present follow-up millimeter observations that demonstrate a periodicity to the phenomenon. While monitoring 3 new periastron encounters, we have detected flares within 17.5\,hours (or 4.6\%) of the orbital phase of the first reported flare and constrained the main emitting region to a stellar height of 3.7--6.8\,R$_\star$. The recorded activity is consistent with the proposed picture for synchrotron emission initiated by a magnetic reconnection event when the two stellar magnetospheres of the highly eccentric ($e$\,=\,0.556) binary are believed to collide near periastron as the stars approach a minimum separation of 8\,R$_\star$\,($\sim$13\,R$_\odot$). The similar light curve decay profiles allow us to estimate an average flare duration of 30\,hours. Assuming one millimeter flare per orbit, DQ Tau could spend approximately 8\% of its 15.8-day orbital period in an elevated flux state. These findings continue to serve as a small caution for millimeter flux points in spectral energy distributions that could contain unrecognized flare contributions. Our analysis of the millimeter emission provides an upper limit of 5\% on the linear polarization. We discuss the extent to which a severely entangled magnetic field structure and Faraday rotation effects are likely to reduce the observed polarization fraction. We also predict that, for the current picture, the stellar magnetospheres must be misaligned at a significant angle or, alternatively, that the topologies of the outer magnetospheres are poorly described by a well-ordered dipole inside a radius of 7\,R$_\star$. Finally, to investigate whether reorganization of the magnetic field during the interaction affects mass accretion, we also present simultaneous optical (VRI) monitoring of the binary, as an established tracer of accretion activity in this system. We find that an accretion event can occur coincident in both time and duration with the synchrotron fallout of a magnetic reconnection event. While the pulsed accretion mechanism has been attributed previously to the dynamical motions of the stars alone, the similarities between the millimeter and optical light curves evoke the possibility of a causal or co-dependent relationship between the magnetospheric and dynamical processes. } 


\maketitle


\section{Introduction}

Pre-main-sequence (PMS) stars are characterized by variability in most wavebands. At millimeter wavelengths, however, the dominant emission process is the optically thin, thermal radiation from the dust located in a circumstellar disk. The integrated flux is therefore a measure of the total amount of cold circumstellar material, which evolves on timescales of roughly 10$^6$\,years. Before the discovery of the first strong millimeter flare toward the young stellar object (YSO) GMR-A in Orion \citep{bower2003,furuya2003}, pronounced short-term millimeter variability was undocumented even though radio variability is well known toward YSOs \citep[e.g.][]{stine1988,white1992}. The lack of millimeter variability studies in the literature is in large part due to the time-consuming nature of observations at these wavelengths, leading to few follow-up measurements. As a result, DQ Tau is only the fourth YSO to be recognized in outburst at millimeter wavelengths. The initial serendipitous detection occurred during an 8-hour observation on 2008 April 19, when the source brightened at 3\,mm (115\,GHz) in a matter of hours to reach a maximum detected flux of 468\,mJy, in comparison to a quiescent value of 17\,mJy in the adjacent days \citep{salter2008}. The flare mechanism was attributed to synchrotron emission from a powerful magnetic reconnection event, probably due to the colliding magnetospheres of the binary components near periastron; and similar in nature to the interacting coronae evidenced toward another millimeter-flaring PMS star \object{V773 Tau A} \citep{massi2002,massi2006,massi2008}. 

DQ Tau ($\alpha_{\rm 2000}$\,=\,04:46:53.06, $\delta_{\rm 2000}$\,=\,$+$17:00:00.1) is a double-lined spectroscopic binary that consists of two relatively equal-mass stars ($\sim$0.65\,M$_\odot$) of similar spectral type (in the range of K7 to M1); as summarized in Table~\ref{parameters}. Its highly eccentric ($e$\,=\,0.556) orbit means that the two stars approach to within 8\,R$_\star$ ($\sim$13\,R$_\odot$) at periastron \citep{mathieu1997}. This makes the system unique in terms of magnetic reconnection events because the minimal separation of the binary is on the order of the theoretical T~Tauri stellar magnetospheric radius of $\sim$5\,R$_\star$ \citep{shu1994,hartmann1994}. This radius is generally defined as the range within which the stellar magnetic field lines remain closed, meaning that the field lines both begin and end at the stellar surface. As a result, the geometry of the system alone predicts an overlap of the two magnetospheres at each periastron event; and as the stars approach one another it may become energetically more stable for the fields to briefly merge together. Field lines rooted on one star break and instantly reconnect with the field lines rooted on the companion, releasing magnetic energy into the surrounding region in the process \citep{vasyliunas1975,hesse1988}.

Magnetic reconnection events like these are still a poorly understood phenomenon. They occur most notably within the solar corona when oppositely-directed field lines are forced together, typically above magnetically active regions where closed magnetic loops, anchored in sunspots, interact with one another \citep{haisch1991}. The associated millimeter emission is explained as the superposition of a gyro-synchrotron spectrum with synchrotron emission \citep{kaufmann1986}, powered by the re-organization of the field lines as they relax into a lower energy state. This solar analogy helps form the basic principles for our stellar scenarios, which can also include magnetic interactions between a star and its accretion disk or a star and a planet forming within its circumstellar disk \citep{phillips1991}. A consistent flare timing near periastron, however, is a strong indicator that both stellar magnetospheres---as the strongest and most stable magnetic structures in the binary system---are contributors to the transient millimeter activity seen toward DQ Tau. 

Here we report on 3 additional flares captured toward this system and discuss what these events mean for the current interpretation of a periodic, star-star magnetic reconnection phenomenon. Our aim is to constrain the nature and regularity of the millimeter emission mechanism by addressing variations in the timing of the flare during the orbit, the duration of the flares, their peak strengths, and the degree of polarization present. For the most recent millimeter flare, X-ray observations were performed in parallel and these will be discussed separately in Getman et al.~(2010, submitted). In addition, periastron events in this system are characterized by variable accretion bursts, most likely driven by the dynamical motions of the binary components \citep{artymowicz1996}. Therefore, a secondary goal of this work is to explore the relative timing of the two interactions, dynamical and magnetospheric, acknowledging that the two mechanisms may exist independently and only share the common periodicity of the binary orbit. To probe the accretion mechanism we present and analyze simultaneous optical photometry measurements, which have been shown to be excellent tracers of the accretion activity in this system \citep[see][]{basri1997,mathieu1997}.


\section{Observations and data reduction}
\label{datared}

\subsection{Millimeter interferometry}

In the follow-up millimeter observations presented here, we chose to observe at slightly lower frequencies in the 3\,mm band (90--95\,GHz) than those specified during the initial discovery (115\,GHz). This strategy takes advantage of greater instrument sensitivities toward lower frequencies and the higher ratio of non-thermal to thermal emission expected toward longer wavelengths. We also supplement the 3\,mm band data with a simultaneous observation in the 1\,mm band (238\,GHz), in order to probe the spectral slope of the transient emission. The technical details for each millimeter data set---all scheduled for observation in the 24-hour period before a periastron event---are summarized in the text below and in Table \ref{mmlog}. While each millimeter facility may respond differently to a high degree of polarization as its linear feeds rotate while tracking the source, we show in Sect.~\ref{polarization} that the polarization fraction is minimal and therefore does not affect the measurements presented here. 


\begin{table}[t!]
\caption{DQ Tau Binary System Parameters \label{parameters} }
\centering
\begin{minipage}[]{0.8\linewidth}
\begin{tabular}{lcc}
\hline
\hline
System Parameter & Value & Source\\
\hline
\noalign{\smallskip}
Stellar Radius [R$_\odot$] & 1.6 & 1 \\
Stellar Mass [M$_\odot$] & 0.65 & 1 \\
Rotation Period [days] & $\sim$3 & 2 \\
Orbital Period [days] & 15.8043 & 1 \\
Orbital Eccentricity & 0.556 & 1 \\
Inclination [$^\circ$] & 157 & 3 \\
Periastron Separation [R$_\odot$] & 13 & 1 \\
Apoastron Separation [R$_\odot$] & 56 & 1 \\
\hline
\noalign{\smallskip}
\end{tabular}
\end{minipage}
\hfill
\begin{minipage}[]{1.0\linewidth}
\begin{tiny}
\textbf{Sources}.~1.~\citet{mathieu1997}; 2.~\citet{basri1997}; 3.~\citet{boden2009}.
\end{tiny}
\end{minipage}
\end{table}


\begin{table*}[tbh!] 
\caption{Millimeter observing log \label{mmlog} } 
\centering 
\renewcommand{\footnoterule}{}  
\begin{tabular}{ccclcccccc} 
\hline 
\hline 
 UT Date & Observatory & Array & Antennas & Baselines & $\nu_{c}$ & Bandwidth & H.A. Range & Optical & X-ray \\
 ~[track start] & ~ & Config. & ~ &  [k$\lambda$] & [GHz] & [GHz] & [hrs] & Data & Data \\
\hline 
 2008 Dec 28 & IRAM & B & 1,\,2,\,4,\,5,\,6 & 26--136 & 90.0 & 1.0 & -2.9 to 5.3 & Yes & No \\
 2009 Mar 17 & IRAM & C & All & 7--53 & 90.0 & 1.0 & -4.1 to 4.1 & No & No \\
 2010 Jan 11 & IRAM & A & All (Ant.~4 fails at 22:30h) & 48--228 & 90.0 & 1.0 & -2.6 to 5.9 & Yes & Yes \\ 
 2010 Jan 12 & CARMA & B & 1,\,2,\,3,\,4,\,5,\,8,\,9,\,10,\,13,\,14 & 36--292 & 92.5 & 3.0 & -3.2 to 5.2 & No & Yes \\
 2010 Jan 12 & SMA & Ext. & All & 27--179 & 238.5 & 4.0 & -3.7 to 4.8 & No & Yes \\
\hline 
\end{tabular} 
\end{table*} 


\textit{IRAM PdBI}.\footnote{IRAM is supported by INSU/CNRS (France), MPG (Germany) and IGN (Spain).} Located in the French Alps, the IRAM Plateau de Bure Interferometer (PdBI) is a 6-element millimeter array capable of measuring two linear polarization directions. We observed DQ Tau at 90\,GHz ($\approx$\,3.3\,mm) on 3 separate UT dates: 2008 December 28-29, 2009 March 17, and 2010 January 11-12. Each observation (or track) was approximately 8\,hours in length and, aside from the array configuration on each UT date, the observations had identical receiver setups of 2$\times$\,1\,GHz (V\,$+$\,H dual-polarization) in full-bandwidth mode (4$\times$\,320\,MHz overlapping quarters). The same standard calibration method was applied to all three tracks: we used the radio source 3C84 for both the passband and flux calibration, and we observed the gain calibrators 0507+179 (a polarized source) and 0446+112 for 2.25 min (45\,sec $\times$\,3 scans) each between every 22.5-min (45\,sec $\times$\,30 scans) on-source observation. All data sets were processed using the GILDAS CLIC and MAPPING reduction software, developed by the Grenoble Astrophysics Group. To extract the flux density, we assumed a model for a point source and we averaged the weighted visibilities for all baselines in 90-sec time steps. The error bars displayed per flux point represent the 1$\sigma$ deviation from the average value. Finally, maps for each track were produced using a natural weighting in the inversion step, and Hogbom cleaning was performed down to the RMS level of the dirty maps. 

\textit{CARMA}.\footnote{Support for CARMA construction was derived from the states of California, Illinois, and Maryland, the James S. McDonnell Foundation, the Gordon and Betty Moore Foundation, the Kenneth T. and Eileen L. Norris Foundation, the University of Chicago, the Associates of the California Institute of Technology, and the National Science Foundation. Ongoing CARMA development and operations are supported by the National Science Foundation under a cooperative agreement, and by the CARMA partner universities.} We observed DQ Tau in the 3\,mm band with the Combined Array for Research in Millimeter-wave Astronomy (CARMA) on UT 2010 January 12. This observation overlapped for 1.5\,hours with the end of the third IRAM track, providing continuous coverage on this date for an 18-hour period during a simultaneous \textit{Chandra} X-ray observing campaign (see Getman et al., submitted). Located in eastern California (USA), CARMA is a heterogeneous interferometer comprised of 23 antennas: six 10.4-m telescopes from the California Institute of Technology/Owens Valley Radio Observatory (OVRO), nine 6.1-m telescopes from the Berkeley-Illinois-Maryland Association (BIMA), and eight 3.5-m telescopes from the University of Chicago Sunyaev-Zel'dovich Array (SZA). Our program was executed while the antennas were being moved out of the CARMA B configuration, and only 10 of the 6-m and 10-m antennas were available to provide baselines between 117 and 946\,m. We used the CARMA Paired Antenna Calibration Systems (C-PACS) to compensate for the rapid phase fluctuations on the long baselines. In the C-PACS observing mode, the CARMA 6-m and 10-m antennas observe the phase calibrator and science target as for normal interferometric observations. Simultaneously, the CARMA 3.5-m antennas observe a bright source at a frequency of 30\,GHz to monitor the atmospheric fluctuations. The observed phase fluctuations at 30\,GHz were then used to calibrate the observed phases toward DQ Tau, and thereby reduce the flux loss due to atmospheric de-correlation on the longer baselines. A detailed description of C-PACS and the data reduction procedures are presented in P\'erez et al.~(2010, in prep). For our observations, the observing cycle consisted of 3-min observations of the gain calibrator 0530$+$135 by both sets of antennas. Then, while the 6-m and 10-m antennas observed DQ Tau for 15 min, the 3.5-m antennas observed the calibrator 0440$+$146. We used 3C273 for the passband flattening and Uranus for flux calibration. The correlator for the 6-m and 10-m antennas was configured with three 500\,MHz bands (or 1.5\,GHz) per sideband that covered the frequencies 89.2--90.7\,GHz in the lower sideband and 94.2--95.7\,GHz in the upper sideband. The data were processed using the MIRIAD data reduction software program, optimized for CARMA. In MIRIAD we used \textit{uvfit} for a point source to determine the flux value and error for time intervals of 5\,min. For mapping we again used a natural weighting for the $u,v$-coverage in the inversion step and cleaning was performed down to 1.5$\sigma$ (where $\sigma$ is the theoretical sensitivity) using the MOSSDI2 package.   

\textit{SMA}.\footnote{The Submillimeter Array is a joint project between the Smithsonian Astrophysical Observatory and the Academia Sinica Institute of Astronomy and Astrophysics and is funded by the Smithsonian Institution and the Academia Sinica.} The Submillimeter Array (SMA) is an 8-element array located on the summit of Mauna Kea in Hawaii. We observed DQ Tau at the higher frequency of 238.5\,GHz ($\approx$\,1.3\,mm), also on 2010 January 12, to achieve an 8-hour overlap with the CARMA track. The array was in its `Extended' configuration, providing 27 baselines spanning 34 to 225\,m. The 4\,GHz bandwidth was uniformly sampled with 128 channels per chunk and a total of 48 chunks. At the start of the night, the opacity was measured independently and the precipitable water vapor (PWV) was estimated to be around 1\,mm, rising to 1.3\,mm at the end of the track. Scans were shortened to 15\,sec while in the extended configuration, and the source was observed for 7.5-min integration loops. The gain calibrators 0423$-$018, 0530$+$135, and 0510$+$180 were observed for 3\,min after every on-source loop. The flux scale was checked every 3\,hours with an observation of Uranus or Mars, and the calibrator 3C273 was used to correct for the passband. The data were reduced using the MIR package for IDL, provided by the Smithsonian Astrophysical Observatory, and subsequently analyzed using the \textit{uvfit} package of the MIRIAD data reduction software program.


\begin{figure*}[htb!]
\centering
\includegraphics[width=17cm]{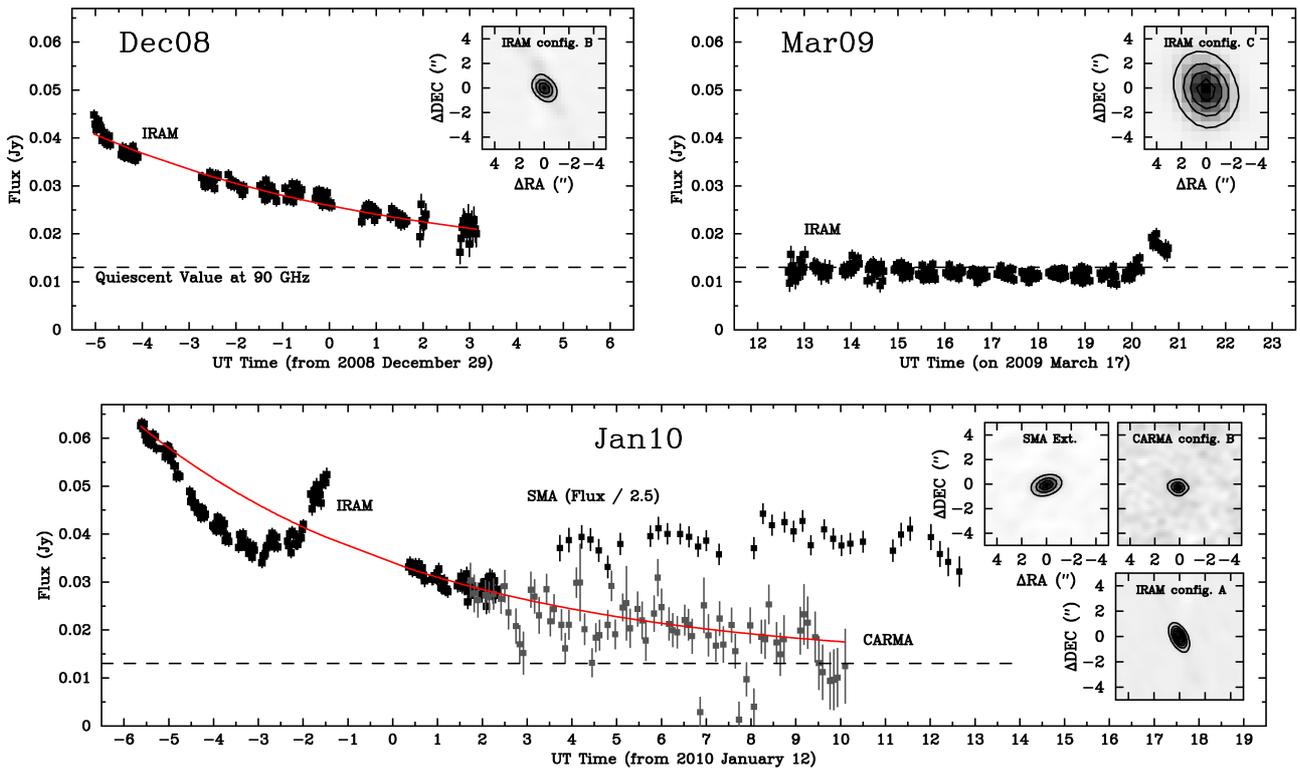}
\caption{The millimeter fluxes versus time for DQ Tau, as observed with the IRAM PdBI, CARMA, and SMA on 2008 December 28-29 (upper left), 2009 March 17 (upper right), and 2010 January 11-12 (lower panel). All fluxes have been determined from a fit to the visibilities using a point source model. Each IRAM data point represents an on-source time interval of 1.5\,min. The CARMA and SMA (divided by 2.5) values are for 5-min intervals. The light curves reveal a (track-averaged) quiescent flux level of 13\,mJy at 90\,GHz during the March IRAM observation and 97\,mJy at 238.5\,GHz for the January SMA observation. During the December and January observations, we caught the flare decay phase and fitted an (identical) exponential decay with an e-folding time of 6.55\,hours (solid red line) to both light curves. We note that the fit does not apply to the initial decay of the original April 2008 light curve (see Sect.~\ref{properties}). The March observation only indicates elevated activity near the end of the track, which we take to be the start of a flare. In the upper right-hand corner of each panel, we give the (unresolved) continuum images for the entire track where the continuum contours are drawn for 3$\sigma$ levels and each $\sigma$ is of the order 1\,mJy\,bm$^{-1}$. \label{mmdatacombined} }
\end{figure*}


\begin{table*}[bht!] 
\caption{Millimeter track and light curve statistics \label{mmstats} } 
\centering 
\renewcommand{\footnoterule}{}  
\begin{tabular}{ccccccccccc} 
\hline 
\hline 
 UT Date & Track ID & Beam Size & $\nu_{c}$ & F$_{start}$ & F$_{end}$ & $\sigma_{avg}$ & Linear & Orbital\,Phase & Hours from & $t$\,=\,0 \\
 ~ & ~ & [$''$] & [GHz] & [mJy] & [mJy] & [mJy] & Polariz. & Coverage [$\Phi$] & Periastron (0.0) & [$\Phi$] \\
\hline
2008 Dec 28 & Dec08-IRAM & 1.34\,$\times$\,1.02 & 90.0 & 43.0 & 20.2 & 1.2 & $<$\,4.65\% & -0.02 to 0.00 & $-$6.8 to 1.5 & -0.056 \\
2009 Mar 17 & Mar09-IRAM & 3.39\,$\times$\,2.79 & 90.0 & 12.6 & 16.9 & 1.1 & $<$\,8.15\% & -0.03 to -0.01 & $-$13.7 to $-$5.3 & -0.007 \\
2010 Jan 11 & Jan10-IRAM & 1.32\,$\times$\,0.70 & 90.0 & 62.4 & 29.2 & 1.1 & $<$\,7.60\% & -0.04 to -0.02 & $-$14.4 to $-$6.5 & -0.067 \\ 
2010 Jan 12 & Jan10-CARMA & 0.87\,$\times$\,0.67 & 92.5 & 30.3 & 12.4 & 3.9 & - & -0.02 to 0.00 & $-$7.2 to 1.1 & -0.067 \\
2010 Jan 12 & Jan10-SMA & 1.25\,$\times$\,1.00 & 238.5 & 95.6 & 85.2 & 5.6 & - & -0.01 to 0.01 & $-$4.7 to 3.6 & -0.067 \\
\hline 
2008 Apr 19 & Apr08-CARMA & 3.66\,$\times$\,1.00 & 113.1 & 468 & 194 & - & - & -0.02 to -0.01 & $-$8.7 to $-$2.7 & -0.020 \\
\hline 
\end{tabular} 
\end{table*} 


\subsection{Optical photometry}
\label{opticaldatareduction}

Optical monitoring of DQ Tau during the first IRAM track on 2008 December 28 was performed simultaneously from the Wellesley College 0.6-m telescope in Wellesley, MA (USA) and the IAC80 0.8-m telescope of Teide Observatory, located in the Canary Islands (Spain). Optical coverage in 2 locations allowed us to monitor DQ Tau throughout the 8-hour millimeter track, and for several hours thereafter. We also obtained simultaneous optical (Teide) and millimeter (IRAM) observations for the final follow-up observation on 2010 January 11. To complete the optical characterization of DQ Tau, additional monitoring was carried out from Wellesley and from Teide Observatory in the one-month period from  December 2008 to January 2009 (covering the weeks before and after the first millimeter follow-up observation). For part of the observations at Wellesley, DMS was a visiting astronomer, while the observations at Teide Observatory were executed in service mode as part of the IAC80 EXTRA observing program for small projects. A complete listing of the optical measurements is provided in Table \ref{optlog} online.

The detector used during the Teide observations was CAMELOT, a
2048$\,{\times}\,$2048 back-illuminated CCD chip with a 0.304$''$ pixel
scale, corresponding to a 10.4$'{\times}$10.4$'$ field of view. The detector on the Wellesley telescope was a 1024$\,{\times}\,$1024 CCD chip with a field of view of 15.6$'{\times}$15.6$'$ and a pixel scale of 1.829$''$ (after binning by 4). We used the standard Johnson V, R, and I filters at both telescopes. All data were reduced using the IRAF data reduction software in the standard way with flats, biases, and for the Wellesley data set, darks. Images with each filter were obtained in blocks of 3 frames. The frames in one block were shifted and co-added. Aperture photometry was performed in IDL using the \textit{cntrd} and \textit{aper} procedures on the co-added images. We used an aperture with an 8-pixel radius for the Wellesley data and a 15-pixel radius for the Teide data.

In order to convert instrumental magnitudes to the standard system, we observed the equatorial standard stars HD 65079 and GD\,50 at different airmasses on 2008 December 25 and 28 at Wellesley. Atmospheric extinction coefficients and zero-magnitude offsets were calculated by comparing our instrumental magnitudes above the atmosphere to the standard values in the literature \citep{menzies1991,landolt1992}. These corrections were then applied for six comparison stars in the vicinity of DQ Tau, observed on the same two nights. The comparison stars are identified in a sample optical image in the online material (see Fig.~\ref{compfig}), and their final absolute magnitudes are provided in Table \ref{comp} online. We estimate a photometric uncertainty of about 0.5\,mag. Two of our comparison stars (\#4 and \#5) were included in the \citet{droege2007} study, and they have a very low Welch-Stetson variability index, confirming that they are sufficiently constant to be used as comparison stars. The absolute magnitudes that we derived for these two comparison stars agree to within 0.2\,mag with the catalog values \citep{droege2007}. Since all six comparison stars are bluer than DQ Tau, we did not attempt to determine a color term, but simply averaged the differential magnitudes obtained with the six comparison stars. Using this method, we estimate that the precision of the repeatability is 0.05\,mag, meaning that we take changes in the light curve larger than the precision value to be real.


\begin{figure}[thb!]
\centering
\includegraphics[width=\columnwidth]{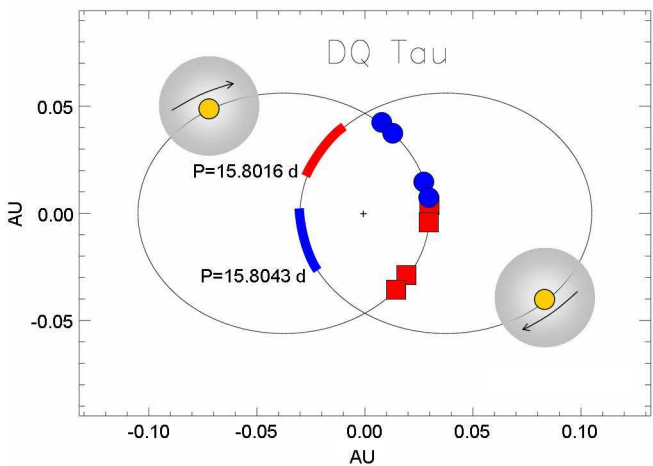}
\caption{The effect of the assumed orbital period on the estimated timing of the flares. A cross indicates the barycenter of the binary system while ellipses trace out the retrograde motions of the binary components. Along one ellipse, we highlight the orbital segment covered by the observations, as determined for each period. On the other ellipse, we indicate the point in the orbit when the trigger event is presumed to have occurred for each periastron monitored. We use blue circles for the \citet{mathieu1997} value of 15.8043\,days and red squares for the 15.8016-day period from \citet{huerta2005}. Set at a greater distance from periastron, we indicate DQ Tau A and DQ Tau B drawn to scale (R$_\star$\,=\,1.6\,R$_\odot$) with an accompanying magnetosphere of R\,=\,5\,R$_\star$. \label{periodfig} }
\end{figure}


\section{Results and analysis}
\label{results}

\subsection{Millimeter flare properties}
\label{properties}

For all millimeter tracks, DQ Tau remains unresolved at all times. Given a distance of 140\,pc to the Taurus star-forming region \citep{kenyon1994}, and a maximum instrument resolution of 0.7$''$ (Table \ref{mmstats}), we can already constrain the millimeter emission to a region within 100\,AU of the binary. The corresponding light curves and track statistics retrieved from the data reduction are plotted in Fig.~\ref{mmdatacombined} and summarized in Table~\ref{mmstats}. In addition, the individual Julian dates and flux values comprising each light curve are provided in Tables~\ref{dec08iram}--\ref{jan10sma} online. 

 
\begin{figure*}[!bth]
\centering
\begin{minipage}[l]{13.5cm}
\includegraphics[width=13.5cm]{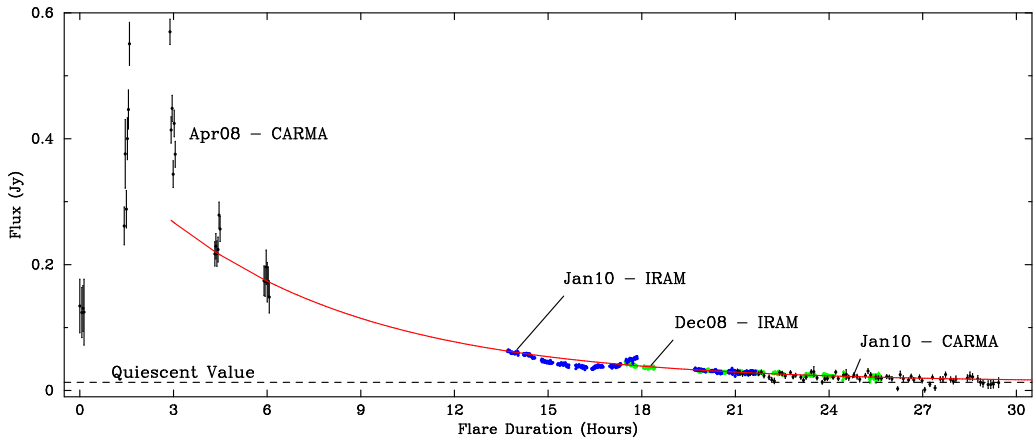}
\end{minipage}\hfill
\begin{minipage}[r]{4.5cm}
\caption{The 3\,mm band light curves are shifted along a common exponential decay curve with an e-folding time of 6.55\,hours. For a zoomed-in version of how well the model fits the decays from Dec08 and Jan10, please refer back to Fig.~\ref{mmdatacombined}. Based on this fit, we loosely establish the upper limit for an average flare duration to be 30 hours. \label{flareduration} }
\vspace{2cm}
\end{minipage}
\end{figure*}


In the follow-up observing campaigns, we report confirmed elevated millimeter activity showing a clear decline in brightness with time on 2008 December 28 and on 2010 January 11 (hereafter abbreviated as the `Dec08' and `Jan10' periastron events). The activity supports a periodic behavior linked to the binary orbit. However, the variable nature of the magnetospheres, and in particular their tendency to fluctuate in size, also means that the exact timing of a flare is understandably difficult to predict. On 2009 March 17 (hereafter the `Mar09' event) the light curve is quiescent until the last hour of the track when it increases in brightness to 1.5 times its quiescent value. At this point the source was nearing an elevation of just 15$^{\circ}$, but since the gain calibrator remains at a reliable elevation, we might only expect a loss in source flux at such a low elevation. Therefore, we conclude that the brightening is a real effect and very likely the start of a flare, occurring a mere 6\,hours later than we had predicted. It is certainly unlikely in a (magnetospheric) collision scenario that a significant flare could have occurred much earlier in the orbit when the stars are at a much greater separation. In this regard, the incompletely-observed onset could raise questions as to whether the millimeter flares share more in common with the optical brightenings, which have been shown to occur for most but not all periastron encounters \citep{mathieu1997}.   
\label{area}

To calculate the timing of each periastron event, we used the orbital parameters determined by \citet{mathieu1997}, specifically JD$_{\rm 0}$\,=\,2\,449\,582.54\,$\pm$\,0.05 and $P$\,=\,15.8043\,$\pm$\,0.0024 days. We note that a revised period has since been published by \citet{huerta2005}, where their more recent data suggests a shortening of $P$ to 15.8016\,$\pm$\,$^{0.002}_{0.006}$. This new period is within the initial error bars of the original, confirming the overall robustness of the orbital period, but the newer value is itself less precise. In Fig.~\ref{periodfig} we illustrate how the assumed period affects the perceived timing of the flares within the orbit. Along one orbital path, we highlight the section that would correspond to our millimeter observations for each proposed period. The original period indicates that our observed events occur as the stars approach one another, whereas the revised period consistently places the observed activity at or after periastron. Due to the irregular nature of the magnetic fields, our data cannot differentiate between the correctness of one period over the other. And, in fact, flares both before and after periastron are possible, as we discuss in Sect.~\ref{timing}. Instead, we suspect that the most energetic events in this system are more likely to occur during the initial interaction, upon approach. Therefore, the orbital phases reported in Table~\ref{mmstats}---and throughout this paper---are calculated using the original period of $P$\,=\,15.8043\,days. 


\begin{figure*}[bht!]
\centering
\includegraphics[width=17cm]{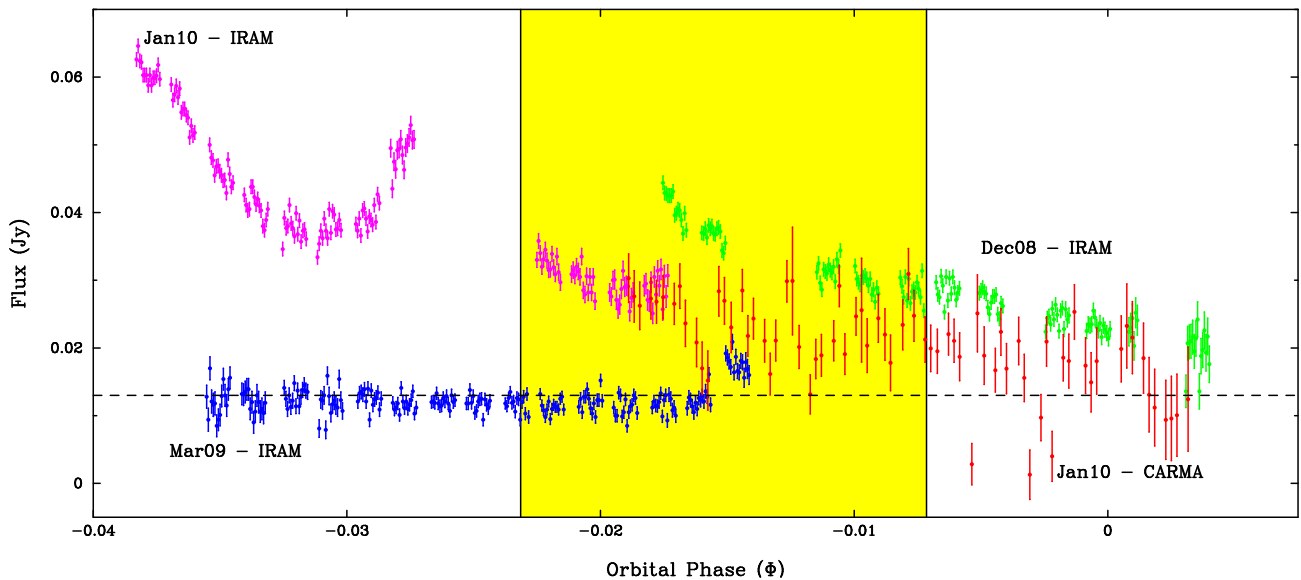}
\caption{A plot of the 3\,mm band fluxes versus orbital phase, illustrating how the activity shifts around slightly in the orbit. An orbital phase $\Phi$ of 0 defines periastron. The shaded region indicates the phase coverage for the initial Apr08 flare and is equivalent to 8\,hours. During the Dec08 and Jan10 follow-up observations, the flare occurred earlier in the orbit, while the Mar09 follow-up observation may have occurred later, but not earlier (at least not within the previous day). \label{phasecover} }
\end{figure*}


In none of the follow-up observations do we succeed in capturing a single flare in its entirety. The longest continuous coverage is for 18\,hours in January 2010 when IRAM, CARMA, and SMA observed DQ Tau in succession. The source remains in an elevated flux state at 90--95\,GHz for the duration of this joint observation, but appears quiescent (within the noise levels) at 238\,GHz. The Jan10-SMA track-averaged flux at 238\,GHz is 97\,mJy versus 13\,mJy at 90\,GHz during the quiescent portion of the Mar09-IRAM track. Together, these flux values define a millimeter spectral index $\alpha$ of $\sim$2, which is consistent with thermal emission from an optically thick disk. During the Jan10 flare, however, the starting flux measurements at SMA (97\,mJy) and CARMA (30\,mJy) document an $\alpha$ of 1.2, providing strong evidence for the presence of non-thermal emission (see Sect.~\ref{axes}).

Since DQ Tau is only above the horizon for 8--10\,hours at any given site, each observation date covers an even smaller window of the recurring activity. However, the separate flare events from Dec08 and Jan10 exhibit very similar flux levels and large-scale decay profiles. When we independently fit an exponential curve to the decaying flux intensities from both dates (shown atop the data in Fig.~\ref{mmdatacombined}), we found identical e-folding times of 6.55\,hours. (We will return to their sub-structure shortly.) If the same e-folding time holds for more (or even all) flares, then each individual observation could loosely correspond to a different portion of the same decay curve. Thus by shifting the separate flare profiles with respect to one another, as is shown in Fig.~\ref{flareduration}, we can attempt to re-construct an ``average'' flare profile. We can even place constraints on the duration of an outburst if we know the peak flux for each flare. With only one peak flux for reference, we extrapolate the exponential fit back through the last flux value from the original Apr08 event. In this way, we estimate a flare duration of approximately 30\,hours. 

This calculation for the flare duration should be regarded with some caution, and taken as an upper limit only in most cases. First, the initial decay slope following the strongest outburst in Apr08 is much steeper than the long steady decays observed in Dec08 and Jan10. As a result, a disjointed superposition of two decay profiles, rooted in two separate energy dissipating processes, may prove a better model and ultimately shorten the flare length. (In Sect.~\ref{decayprocesses} we discuss these loss mechanisms further.) Secondly, the assumption of a similar peak flare intensity from one periastron to the next is one property where we lack sufficient millimeter data. To reinforce this idea, simultaneous X-ray observations during the Jan10 event indicate only a relatively weak X-ray flare; one that is hardly powerful enough to expect a radio counterpart with a peak flux as explosive as the original Apr08 event (Getman et al., submitted). Finally, it is improbable that the peak fluxes are identical given the complicated nature of the magnetic fields, the dependence on the ionized particle reservoir, the magnetic energy released during an event, as well as the diversity that we can already document in the decay profiles. However, this method provides some necessary constraints to characterize the millimeter activity. In turn, for a 15.8043-day period assuming one 30-hour flare per orbit, we suspect that DQ Tau could spend up to 8\% of its time exhibiting excess flux that is unrelated to the thermal continuum emission from its circumbinary disk. 

If we take this analysis one step further, we can use the relative offsets from Fig.~\ref{flareduration} (e.g.~$\sim$13.5\,hours for Jan10 and $\sim$17.5\,hours for Dec08) to estimate the orbital phase $\Phi$ for the initial outburst event (or $t$\,=\,$0$). These are the orbital points indicated along the second ellipse in Fig.~\ref{periodfig}, and in numerical form in Table \ref{mmstats} (Column 10) where $\Phi$\,=\,0.0,1.0 defines periastron. If we now continue to assume that the Mar09 track signals the start of a large flare, then the Jan10 and Mar09 flares exhibit the largest separation in orbital phase, suggesting an outburst event window of $\sim$22\,hours (or $\sim$6\% of the orbital phase). Physically, this window encompasses a stellar separation of 8--13\,R$_\star$ and is equivalent to roughly one-third the time needed for the stars to swap positions about the system center ($\sim$2.8\,days; Basri et al.~2010). In Fig.~\ref{phasecover} we plot the millimeter flux versus the orbital phase for all tracks, to show how the elevated activity shifts back and forth in the orbit, yet consistently occurs in the day before periastron ($-$0.06\,$<$\,$\Phi$\,$<$\,0.0). Together, the estimated flare duration plus the variable orbital timing for the outburst event, define a window of breadth 52\,hours (or $\sim$2.2 days) when the millimeter flux may exceed the thermal quiescent value.  
\label{sepdist}

While the similarity in the large-scale decay times for both Dec08 and Jan10 is quite striking, the curves show sub-structure that is rather diverse. During the Apr08 and Dec08 events, we captured a smooth exponential decay, exhibiting only very small variations in the measured flux values. The Jan10 light curve, however, features a sharp break in the general decay profile one hour into the observation, initially dropping more rapidly before giving rise to a milder secondary peak, and then finally resuming the original decay profile. (We note that the 2-hour gap in the data during this time was the result of an antenna error, and that the observations quickly resumed without the faulty antenna.) We interpret the sub-structure from Jan10 to be due to a secondary event, occurring about 15\,hours after the estimate for the initial outburst, and involving a smaller energy release than the preceding flare. In fact, this occurrence of successive events suggests that a series of (probably less powerful) flares can take place, possibly in lieu of one giant outburst. 

The Jan10 multiple-peak observation, which occurred earliest in the orbit, was presumably less powerful (drawing on the X-ray analysis) than the original Apr08 outburst, which occurred closer to periastron (Fig.~\ref{phasecover}). While we lack information on the full millimeter light curves to make a conclusive statement, we note that a similar inverse relationship between flare intensity and peak timing was documented toward V773 Tau A \citep{massi2008}. One caveat to the assumption of a lower peak flux is that it shortens our estimate for the flare duration, and thus the phase of $t$\,=\,0. On the other hand, a superposition of multiple flares spaced slightly apart in time can lead to both shortened or extended periods of activity, which can only be better characterized by a larger monitoring program. 

Additional small-scale variations in the total flux density are likely due to atmospheric and instrumental effects. Larger systematic changes can help us test for strong linear polarization using the new, dual linear polarization receivers available at IRAM PdBI in combination with Earth rotation polarimetry, as described in \citet{trippe2010}. There the authors show how the difference in flux between the two orthogonal polarization feeds, divided by their sum, changes in a systematic way when a linearly polarized source transits. Unfortunately, this technique and the instrumentation available does not allow us to constrain the presence or amount of circularly polarized light. Using this method in the absence of detecting all 4 Stokes parameters, we determine 3$\sigma$ upper limits of 4.65\% and 7.60\% for the linear polarization during the flare decay phases from Dec08 and Jan10, respectively. During the quiescent Mar09 track, our upper limit is 8.15\%. In Sect.~\ref{polarization}, we discuss the implications of the linear polarization fractions for the proposed picture and the emission mechanism.

\subsection{Coincident optical brightenings}
\label{optbright}

DQ Tau is perhaps best known as the first system to be studied in terms of pulsed accretion flows to explain its optical variability near periastron. The binary has therefore been characterized extensively at optical and near-infrared wavelengths, using both photometric and spectroscopic observations \citep{herbst1994,mathieu1997,basri1997,huerta2005,boden2009}. In Table~\ref{optlog} online, we provide the photometry results from our own observing campaigns in December 2008, January 2009, and January 2010. We note here that when we compare the Teide and the Wellesley data obtained on 2008 December 28---the only night for which the two telescope observations overlap in time---we find a slight offset between the magnitudes obtained with the two telescopes, which is within the uncertainties stated in Sect.~\ref{opticaldatareduction}. Consistency can be achieved by subtracting 0.08\,mag from the Wellesley V magnitudes and 0.05\,mag from the Wellesley R and I magnitudes. We emphasize that, while the online photometry table does contain the original \textit{un}shifted magnitudes per telescope, we have shifted these data accordingly for all plots displayed here for the purpose of our analysis and the discussion that follows. 
 
\begin{figure}[b!]
\includegraphics[angle=90,width=9.2cm]{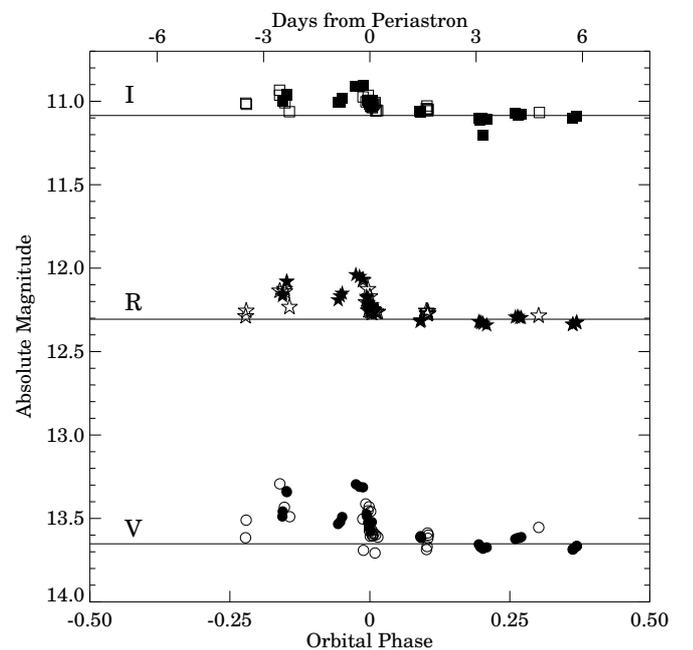}
\caption{The VRI photometry values wrapped with the 15.8043-day orbital period for DQ Tau, where $\Phi$\,=\,$\pm$0.5 indicates apoastron and $\Phi$\,=\,0 is periastron. The optical brightenings occur within a 5-day window near periastron, as was first shown by \citet{mathieu1997}. Filled symbols represent data taken with the Teide IAC-80 telescope and unfilled data points are from the Wellesley telescope. Horizontal lines indicate the quiescent absolute magnitude per filter. \label{dqtauopticalwrap1} }
\end{figure}

\begin{figure}[bh!]
\includegraphics[angle=90,width=9.2cm]{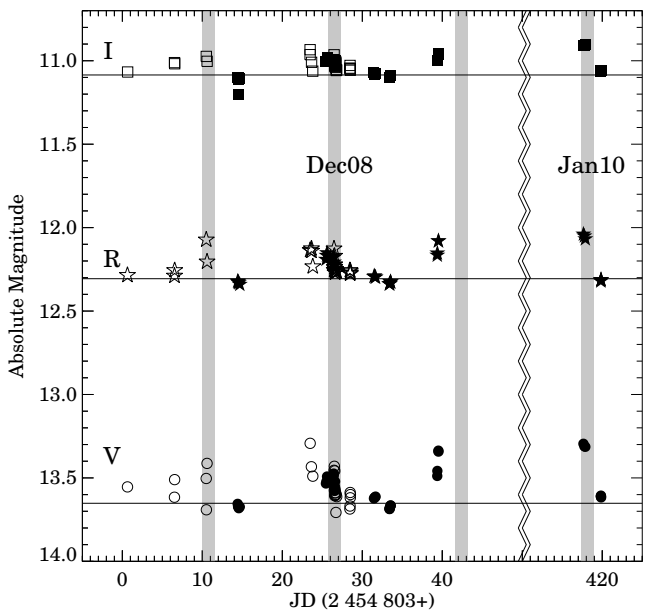}
\caption{A plot of the VRI photometry values for the Julian dates from December 2008 through January 2009, as well as January 2010. The shaded columns indicate periastron events. The source brightens irregularly prior to all four newly documented periastron encounters. \label{dqtauopticalwrap} }
\end{figure}

In Fig.~\ref{dqtauopticalwrap1} we plot our photometry values first as a function of the orbital phase $\Phi$ for all three optical filters. In the optical V band, the source is known to brighten by $\approx$0.5\,mag within an orbital phase window of breadth 0.30\,(or $\sim$5\,days), centered on a phase of $-$0.1 (or about 1.6\,days before closest approach). The V filter always shows the greatest amplitude change, but the variations are mirrored in all filters. This ``bluening'' effect caused by increased veiling is just one of the expected results from the ongoing accretion processes \citep{basri1997}. We also observe the same 5-day activity window for the optical brightenings, which is a window 2.3 times broader than the millimeter window documented thus far. Our photometry results are indeed consistent with previous studies \citep[see][Fig.~4]{mathieu1997}. One unique feature in our data, however, is the apparent double-peaked nature of these most recent brightenings, indicating a clustering of events around two phases. We caution against over-interpreting this effect since the data set is sparsely-sampled, covers only 4 periastron encounters, and is heavily dominated by the Dec08 event in particular (see Fig.~\ref{dqtauopticalwrap}). Nevertheless, the behavior is intriguing given the emerging picture for star-star reconnection events, and we will return breifly to this result in Sect.~\ref{link}.


In Fig.~\ref{dqtauopticalwrap} we unwrap the optical data and show the results of continuous monitoring from December 2008 to January 2009, as well as several measurements from January 2010. We indicate the absolute magnitude versus Julian Day, where the bi-weekly periastron events are indicated with a shaded column. The source is found to increase in brightness in the days leading up to each of the 4 periastron encounters that we monitored, even though optical brightenings were not seen for all encounters monitored by \citet{mathieu1997}. Of particular interest are the second and fourth shaded columns in Fig.~\ref{dqtauopticalwrap} (at JD\,=\,2\,454\,829 and 2\,455\,221, respectively), which represent the dates for the Dec08 and Jan10 observations, confirming coincident optical and millimeter activity on each date. No optical data is available for Mar09 and Apr08, corresponding to a time of year when the source is below the horizon at night.

\begin{figure}[t!]
\centering
\includegraphics[width=7cm]{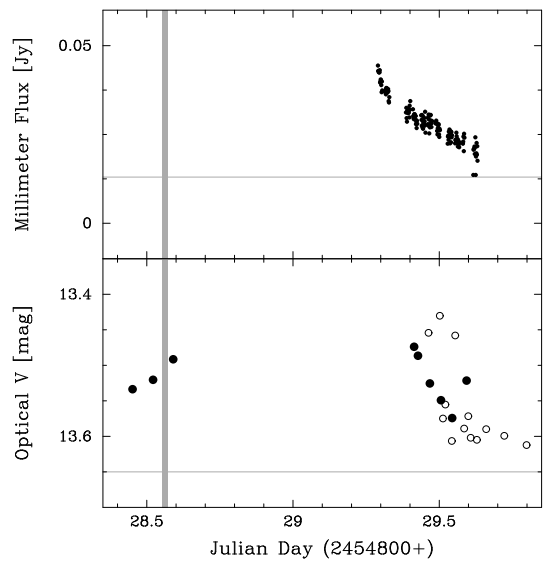}
\caption{A plot showing the simultaneous millimeter (upper panel) and optical (lower panel) coverage for the Dec08 periastron encounter. A thin gray horizontal line in each panel represents the quiescent value. The optical light curve begins to brighten the night before, returning to quiescence in time with the millimeter decay. We indicate with a thick vertical line $t$\,=\,0 for the estimated time of the trigger event, assuming an average 30-hour flare. In this case, the optical activity appears coincident in both time and duration with the millimeter activity. }
\label{timestats}
\end{figure}

The Dec08 event offers the most extensive simultaneous optical coverage. In Fig.~\ref{timestats} we show how the optical and millimeter outbursts are related in time. On the night before the millimeter track, the optical light curve begins to brighten. During the millimeter decay the following night, the optical light curve decreases in step with the millimeter flare. This brightening and subsequent decay shows that the optical activity is not only coincident in time, but also in duration, with the average millimeter profile.


\section{Discussion}

\subsection{Signatures of synchrotron emission}

The millimeter emission is almost certainly non-thermal in origin given the sudden flare onset and the fast decay time. In the solar analogy, it is generally accepted that this type of extraneous emission from high-energy electrons is synchrotron in origin, and that emission at shorter millimeter wavelengths results from the most energetic (MeV) electrons \citep{silva2007}. We eliminate non-relativistic cyclotron radiation as a primary emission process because the magnetic field strength required to produce cyclotron emission ($B=2 \pi\,m_e / q$) at 90\,GHz is 30\,kG; increasing to 80\,kG at 238\,GHz. These values are much larger than the 1--6\,kG surface field strengths expected (or measured) for TTSs \citep{guenther1999,johnskrull2007}. Instead, the electron population probed by our observations must possess mildly to highly relativistic properties, representing gyro-synchrotron and synchrotron radiation mechanisms, respectively. 

\label{axes}
In this subsection, we follow closely an analysis similar to that presented in \citet{massi2006} for synchrotron emission from the interacting coronae of V773 Tau A. We model the outer (large-scale) structure of the DQ Tau magnetospheres to be well-ordered dipoles and we take the magnetic axes to be aligned parallel to the rotational axes, assumptions that are consistent with recent 3D extrapolations of the coronal field topology for several young stars \citep{gregory2008}. We do assume that the rotational axes are both aligned perpendicular to the plane of the system, although we acknowledge that, in rare cases, large misalignments have been documented around a short-period binary \citep{albrecht2009}. We know that the DQ Tau system is inclined 157$^{\circ}$ to the line-of-sight \citep{boden2009}, and we presume that the reconnection events occur in the equatorial region between the two stars. For a synchrotron spectrum, \citet{ginzburg1965} showed that the maximum spectral frequency (in Hz) is proportional to the perpendicular component of the magnetic field $B_{\perp}$ (in Gauss) and the Lorentz factor $\gamma$ squared:
\begin{equation}
\label{vmaxgeneral}
\nu_{\rm max} = 1.2 \times 10^6 B_{\perp} \gamma^{2}
\end{equation}
\noindent Since millimeter brightenings have been consistently observed near 90\,GHz, we conservatively assume this value for $\nu_{max}$, meaning that there must be an electron population whose relativistic properties in vacuum satisfy the following relation:
\begin{equation}
\label{vmax}
B \gamma^{2} = 8.1 \times 10^4
\end{equation}
\noindent Using an average surface field strength of 3\,kG, we can calculate the Lorentz factor for several relevant distances, including: half the minimum binary separation ($d$\,=\,4\,R$_\star$, $B$\,=\,47\,G, $\gamma$\,=\,42); the minimum binary separation ($d$\,=\,8\,R$_\star$, $B$\,=\,6\,G, $\gamma$\,=\,116); and the distance to the inner rim of the circumbinary disk in the case of a star-disk interaction ($d$\,=\,54\,R$_\star$, $B$\,=\,0.02\,G, $\gamma$\,=\,2012). All scenarios give rise to highly relativistic ($\gamma$\,$\gg$\,1) particles, and thus synchrotron emission. For gyro-synchrotron ($\gamma$\,$\approx$\,5) emission only, contributions would be restricted to stellar heights of 0.3--1.5\,R$_\star$ and most likely indicate single-star magnetic activity that is independent of the orbital period.

Due to the nature of synchrotron emission and the local magnetized environment, competition between radiation and collisional losses represent the largest impactors on the exponential decays that characterize our millimeter light curves. We start with the following expression for synchrotron radiation losses (in hours) for a well-ordered field \citep{blumenthal1970}:
\begin{equation}
\label{tsync}
\tau_{s} = \frac{1.6 \times 10^{5}}{B^{2} \gamma}
\end{equation}
\noindent In combination with Eq.~\ref{vmax}, we can solve for the two unknowns when we use a synchrotron decay time $\tau_{s}$ equal to the e-folding time of 6.55\,hours. We find $B$\,=\,19\,G and $\gamma$\,=\,65. For the full range of dipolar field strengths expected for a TTS (1--6\,kG), this result localizes the main emitting region to a stellar height of 3.7--6.8\,R$_\star$. We note that the values are centered around the theoretical size for a T~Tauri magnetosphere. They are also consistent with a site located halfway between the two stars for the separation distances at the times of outburst ($\sim$8--13\,R$_\star$ as determined in Sect.~\ref{sepdist}). This range compares well with the coronae loop size infered by Getman et al.~(2010, submitted) to explain the X-ray activity, which derives from a related process but a separate electron population. Alternately, if significant emission were to be observed near a $v_{\rm max}$ of 238\,GHz, then the stellar height range identified would increase (unless a weaker stellar magnetic field or a faster decay time at the higher frequency were to compensate).

Next we constrain the maximum density of the electrons spiraling along the field lines by calculating what the thermal Coloumb collisional losses (in hours) would need to be in order to shorten the observed decay time. We use \citep{petrosian1985,massi2006}:
\begin{equation}
\label{tcoll}
\tau_{c} = 4.16 \times 10^{8} \frac{\gamma}{n_{e}}
\end{equation}
\noindent to derive a maximum electron density of $n_{e}$\,$\le$\,3.7\,$\times$\,10$^{9}$\,cm$^{-3}$. A final check of the condition \citep{ginzburg1965}:
\begin{equation}
\label{vacuumcondition}
\nu \gg \nu_{c} \simeq 20 \frac{n_{e}}{B_{\perp}}
\end{equation}
\noindent confirms that the vacuum approximation used throughout these calculations is valid in the case of the DQ Tau magnetospheres as accelerated electrons spiral down from large stellar heights.

The fact that synchrotron losses explain the light curve profiles well implies that the relativistic electron reservoir is sufficiently confined, or trapped, within the global magnetic structure. One of the key conclusions from the \citet{massi2006} model for V773 Tau A was that electrons must leak out at magnetic mirror points in order to diffuse the synchrotron emission fast enough to achieve their observed e-folding time of 2.31\,hours. While evoking this effect is unnecessary late in our own flare timeline, we note that a much steeper decay is present in the initial hours of the Apr08 flare (refer to Fig.~\ref{flareduration}). It is possible that a similar leakage of electrons could also occur in the DQ Tau system shortly after the initial outburst, or perhaps as (accreting) charged particles are initially expelled from the system in a manner similar to a coronal mass ejection in the Sun. Or perhaps a more probable scenario is that the Apr08 decay, like the Jan10 dip, is the result of the Neupert effect and the thick-target mechanism (see Getman et al., submitted). We do rule out several geometrical effects, including an eclipse due to rotation, because the stellar rotational periods have been determined to be $\sim$3\,days \citep[see][]{basri1997}, while the maximum deviation from the decay profile lasts at most 5.5\,hours. In the end, the geometry-independent, natural synchrotron decay process (vs. particle loss or leakage) as the primary loss mechanism agrees much better with the persistent, large-scale decay profiles observed from one flare to the next. 
\label{decayprocesses}

\subsection{De-polarization effects}
\label{polarization}

Synchrotron emission predicts a high fraction of linearly polarized light, up to 70\% in a well-ordered magnetic field, although it is rarely observed to be maximally polarized \citep[e.g.][]{sokoloff1998,trippe2010}. Our own observations limit the polarization fraction toward DQ Tau to $<$\,5--8\%. The measurements sample both a quiescent state as well as two flare decays, and do not rule out the presence of linear polarization altogether. Instead, we explore de-polarization effects that can reduce the observed polarization fraction.

Polarized light originates in regions where the magnetic field lines are stable, well-ordered, of a single orientation, and of a single magnetic polarity. Since the plane of polarization is perpendicular to the magnetic field, a tangled field unresolved by the telescope beam leads to beam averaging of the different polarization vectors, effectively de-polarizing the light. In Sect.~\ref{area} our resolution constrained the emission to a region 100\,AU from the binary, an area that encompasses the two independent magnetospheres. Within this view, dipolar magnetospheric field lines on one star begin in its northern magnetic hemisphere and end in its southern magnetic hemisphere, effectively doubling back 180$^{\circ}$ in direction within the telescope beam. In addition, the field polarity of one star may be reversed with respect to the companion's. In fact, we show in Sect.~\ref{alignment} how this is probably the case for DQ Tau. Therefore, the geometry of the field lines where the accelerated electrons are trapped has a very important impact on the net polarization observed, and any positive polarization detections may favor emitting regions at larger stellar heights where the magnetic field is simpler and well-ordered, since higher-order field components typically fall off more quickly \citep{gregory2008}.

Nevertheless, linear polarization has been measured toward V773 Tau A, the only other known PMS binary to exhibit evidence of star-star magnetic interactions. \citet{phillips1996} detected a significant fractional polarization of 2\% at cm wavelengths. However, follow-up polarization observations of V773 Tau A during periods of millimeter activity produced only non-detections at both millimeter and centimeter wavelengths \citep{massi2008}. Generally, polarization is considerably weakened toward the radio end of the spectrum, mainly due to Faraday rotation effects \citep{ginzburg1965}. The effect occurs as linearly polarized light passes through a dense, magnetized medium causing the polarization angle $\theta$ to rotate by an angle equal to $RM$\,$\lambda^{2}$ where $RM$ is the rotation measure (in rad\,m$^{-2}$) defined as:
\begin{equation}
\label{farrot}
RM = 8.1 \times 10^{5} \int n_{e}  B_{||} \, ds
\end{equation}
\noindent Here $n_{e}$ is the electron density (in cm$^{-3}$), $B$ is the longitudinal magnetic field strength (in Gauss), and $ds$ is the line-of-sight path length (in pc) through the magnetized medium. When the rotation measure is larger than $\sim$10$^{5}$\,rad\,m$^{-2}$ at 3.3\,mm, Faraday de-polarization is complete in a homogeneous medium. This is equivalent to a rotation angle of 90$^{\circ}$, although rotation angles as small as 5--10$^{\circ}$ (or $RM$\,$\approx$\,10$^{4}$) can also be sufficiently effective, depending on the method detection threshold and the intrinsic polarization fraction.

Faraday rotation can occur in any magnetized region along the line-of-sight, including the interstellar medium and the Earth's ionosphere. However, the synchrotron emission from DQ Tau predicts large electron densities (10$^{9}$\,cm$^{-3}$) and strong magnetic fields (3 orders of magnitude greater than the Sun's field), suggesting that the emitting region itself has the potential for the greatest de-polarization effect. To test this possibility, we model electrons trapped in the magnetospheres with a shell of outer radius 5\,R$_\star$ and a thickness of 1\,R$_\star$. The electron density is set to the maximum determined from synchrotron losses and taken to be constant throughout the shell. The magnetic field strength is modeled as a dipole with a surface field strength of 3\,kG. In this scenario, we derive an $RM$ of $\sim$\,10$^{9}$\,rad\,m$^{-2}$, which undoubtably results in complete de-polarization of the synchrotron emission. 

Conceding that this initial calculation may represent a case of maximized extremes, we can do the same calculation for an electron density of 10$^{3}$\,cm$^{-3}$ (essentially an upper limit in a molecular cloud of typically 10$^{3}$ \textit{molecules} per cm$^{-3}$), a shell thickness of 0.1\,R$_\star$ (equivalently the size of the stellar corona), and a reduced dipole field of surface strength 1\,kG. We find an $RM$ value of $\sim$10$^{2}$, which is below the cutoff for complete de-polarization. To recover a factor 100 in the $RM$, we determine that a minimum density of 10$^{5}$--10$^{6}$\,cm$^{-3}$ is needed to sufficiently de-polarize the millimeter emission. These values are representative of the density in the uppermost layers of the circumbinary disk, and for a source experiencing (simultaneous) accretion, should be easily obtained, at least along the accretion streams. Therefore, we predict that the synchrotron source is sufficiently self de-polarizing to result in the complete absence of linear polarization for most geometries.

\subsection{Orientation and topology of the magnetospheres}

\begin{figure}[tb!]
\centering
\includegraphics[width=8.5cm]{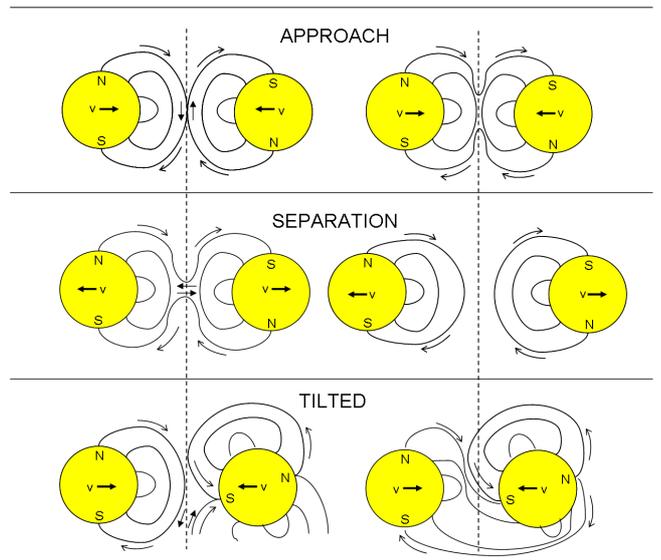}
\caption{An illustration of the scenarios for reconnection events in the DQ Tau binary system during approach (upper panel) and separation (middle panel). In the lower panel, we show a variation to the upper panel if the magnetic axes are tilted with respect to one another. This misalignment can also occur into or out of the page, causing field lines to cross at intermediate angles that may or may not result in reconnection. Similar variations to the middle panel are also possible. \label{reconnectionscenarios} }
\end{figure}

A dipole representation of the magnetospheres remains an adequate and consistent model for the analysis, and is the simplest valid structure. When we illustrate the corresponding field lines in Fig.~\ref{reconnectionscenarios}, there are, in fact, two scenarios for the timing of the reconnection events: during both the approach and separation phases. During approach, the field lines are oppositely-directed in the equatorial region and undergo compression as the stars approach periastron and the fields repel one another. At the vertical boundary layer between the two magnetospheres, where their merging plasma flows and induced electrical currents resist one another, reconnection can occur. The two magnetospheres join together via shared field lines that begin on one star and now end on the other. During separation, the global field is split into two closed magnetospheres through stretching, and thus compression of the lines near the orbital plane of the system. Both scenarios release energy into the surrounding region, accelerating charged particles at large stellar heights down along the magnetospheric field lines toward the star, which is when both scenarios produce (gyro-)synchrotron radiation in a relatively indistinguishable manner. 
\label{timing}

In the figure, we point out that the magnetic axes are drawn with an inverse alignment. In an aligned system, trying to connect the field lines from the magnetic north of one star to the magnetic south of the other results in crossed lines, which would quickly reconnect, reverting back to two dipole structures. Thus, in this arrangement, the global magnetic topology is strictly maintained. The magnetic density still increases upon approach, building up magnetic energy stores, but the boundary layer between the two magnetospheres is imperceptable to the local plasma and electrical currents. There is far less resistance to the re-arrangement (e.g.~compression) of the field lines, as compared to the reverse case, and reconnection is not favored as a result. 

Assuming parallel rotational axes perpendicular to the orbital plane, and that magnetic axes tend to be aligned with the rotational axis in many astrophysical bodies (as we justified in Sect.~\ref{axes}), the two magnetospheres should be either aligned (0$^{\circ}$) or inversely aligned (180$^{\circ}$). Undoubtedly, in this bimodal interpretation, the DQ Tau magnetospheres therefore must be inversely aligned to produce the flares observed. This does present an interesting consequence for the millimeter flares if one of the magnetic fields were to flip, as the Sun's field is prone to do once every 11 years; an effect that has been observed toward other PMS stars with short rotation periods \citep{donati2008a,fares2009,petit2009}. In this case, if the magnetospheres are the principle mechanism for the flaring phenomenon, and if their magnetic axes are indeed more or less aligned, then we could expect on-and-off periods of millimeter activity near periastron.

Finally, there are the oblique cases when the field lines forced together are not similary- or oppositely-directed, but rather cross at an intermediate angle. This can occur if the magnetic axes are tilted with respect to one another. The occurrence of reconnection then depends on the orientation angle, the field strengths, the merging systems of flux, and the resistivity to the topological changes caused by the induced currents. Consequently, one conclusion to make is that the DQ Tau magnetospheres must be misaligned by a significant angle to produce reconnection events and their associated millimeter flares.
\label{alignment}


The other possibility is that the magnetospheres are poorly modeled by a dipolar field between 4--7\,R$_\star$ (or 6.4--12\,R$_{\odot}$), which is the half-distance separation of the two stars during the observed flares. In the Sun, the dipole component dominates beyond 2.5\,R$_{\odot}$, but in PMS stars the relative strengths of the field components can vary widely \citep{gregory2008}. For example, toward the TTS V2129 Oph the octopole component was found to dominate out to 6.7\,R$_\star$ \citep{donati2007}, providing ample opportunity in a binary system for opposite-polarity regions to interact and reconnect at large separations. Other deviations include the winding up of the field lines as the stars rotate faster in the equatorial region. This effect is amplified with respect to the Sun given that these stars rotate 10 times faster. However, we expect that the winding effect should act to slow the rotation in an aligned system, and encourage reconnection as usual in an inversely aligned system. 

The accretion streams can also produce local changes in the field lines, including compression and ordering of the magnetosphere during episodic events. It is unclear what should be expected from a collision of two streams with equivalent mass accretion rates, but it may mirror the complex and unstable merging of two stellar jets, as modeled by \citet{mundt2010}. There the authors predict that, as the outflows in a binary system combine at large stellar distances, reconnection events should result near the intersection point. Finally, the trigger might simply be a inter-coronal interaction very similar to the \citet{massi2006,massi2008} scenario for V773 Tau A. There the extended magnetic structure was imaged at a stellar height of 48\,R$_{\odot}$. Thus, if similarly extended structure were present on the stars in the DQ Tau system, then flares could result at any time in the orbit. Although currently, all observations outside our defined outburst event window report quiescent fluxes (Salter et al.~2008; Guilloteau et al., in prep), suggesting that this is not the most robust description for the current observations. Of course, we cannot rule out millimeter contributions from a combination of scenarios.


\subsection{Optical emission mechanisms}
\label{link}

The coincident timing and duration of the optical and millimeter brightenings in Fig.~\ref{timestats} is intriguing, but the physical interpretation remains a challenge. While we have presented evidence that the millimeter emission arises from recurring star-star reconnection events, the origin of the optical emission can be rooted in both dynamical processes (e.g.~accretion pulses due to the binary motions) and magnetospheric processes (e.g.~reconnection events). In this section, we consider whether both processes are required to explain the optical behavior. 

Flare models often include an optical emission component due to the heating and ionization of the chromospheric plasma where the non-thermal electron population spirals down into the stellar atmosphere, typically at a magnetic field footprint \citep{haisch1991,gudel2002}. In this picture, an optical brightening is normally expected to accompany the millimeter activity, just like we currently observe. It is thus tempting to recall the double-peaked nature of the optical brightenings in Fig.~\ref{dqtauopticalwrap1}, in light of our scenario for interacting magnetospheres, which predicts two reconnection events associated with first the joining and then the separation of the magnetospheres. The data presented in \citet{mathieu1997} do not preclude optical brightenings clustered around two separate phases, spaced apart in time by roughly 3\,days ($\Delta$$\Phi$\,$\approx$\,0.2). In addition, \citet{mathieu1997} also determined that enough magnetic energy is available in the outer magnetospheres to power the average DQ Tau optical brightening, even though the authors questioned whether the field could regenerate itself on a bi-weekly basis. Although our millimeter observations do not cover back-to-back periastron encounters, our consistent flare detection rate suggests that reconnection occurs during most, if not all, orbits. However, it remains unclear whether this result favors a quick and efficient regeneration of the field or an alternative mechanism to drive the optical brightenings. 

Instead, other observational results favor optical brightenings linked to independent accretion processes. \citet{mathieu1997} noted first that the optical brightenings occur quite far from periastron to be the result of reconnection, as well as the spectral bluenings seem to disagree with the optical flat-spectrum (or ``white-light") continuum enhancements that more typically accompany flares \citep[e.g.][]{hudson1992,gudel2004}. Likewise, some of the DQ Tau optical brightenings appear to last continuously for up to 4\,days \citep[Fig.~2 of][]{mathieu1997}, much longer than an ``average" millimeter flare. In Sect.~\ref{optbright} we determined that the observed millimeter activity window was 2.3 times smaller than the optical, but we concede that our phase sampling was intentionally restricted to the 24-hour period before periastron, and therefore we lack a detailed overview of the millimeter perspective at larger separations. This fact, combined with the \citet{massi2008} evidence for interacting coronae at large ($\ge$30\,R$_\odot$) stellar separations, suggest that only a prolonged multi-wavelength monitoring program over several cycles, and covering all orbital phases, can best characterize the relationship between the optical and millimeter activity, and the relevant dynamical and magnetospheric contributions. 

In summary, it is not straight-forward to assign all of the elevated optical acitivity in this sytem to dynamically induced pulsed accretion events, particularly near periastron. Accretion appears to be ongoing at different levels throughout the orbit \citep{basri1997}, with features that fit well the models by \citet{artymowicz1996}. Reconnection events, on the other hand, have only been documented near periastron thus far, but seem to always accompany an optical event. Thus, in all likelihood, reconnection and accretion processes are both contributing to the elevated optical activity, occurring simultaneously at periastron. To better ascertain the causal or co-dependent nature of the dynamical and magnetospheric processes, it is necessary to separate out the optical effects of each process.


\section{Conclusions}

We confirm periodic, elevated millimeter flux levels toward the DQ Tau PMS binary while monitoring 3 recent periastron encounters. The regularity of the flare timing is consistent with the proposed scenario for colliding magnetospheres within a day of closest approach, although the initiating event appears to vary within a 22-hour window corresponding to a binary separation of 8--13\,R$_\star$. The flaring mechanism can be explained by synchrotron emission from highly relativistic electrons accelerated near the reconnection site that begin to spiral down along the magnetospheres toward one or both stars. The main emission region is localized near a stellar height of 3.7--6.8\,R$_\star$, about halfway between the two stars at the time of an event. Synchrotron losses easily explain the similar (late-time) decay profiles, indicating both a well-confined electron population and a large stable magnetic structure from one event to the next. We estimate that the flares can last up to 30\,hours per event, corresponding to 8\% of the orbital period. In addition, multiple millimeter flares during a single periastron event have been observed within approximately 15\,hours ($\Delta$$\Phi$\,$\approx$\,0.04) of one another, with the secondary event being less powerful. The succession of events may correspond to first the merging and then the separation of the two magnetospheres, or simply to a trade-off between a slow sequential release of stored magnetic energy in lieu of one large outburst. A star-star magnetic reconnection event remains the simplest, most straight-forward interpretation in terms of the timing of the activity, the regularity of the occurrence, the magnetic field strengths implied, and the field sizes.

We measure an upper limit of 5--8\% for the fractional linear polarization of the light, but we predict that both beam dilution and Faraday rotation in the DQ Tau system are probably sufficient to result in a net polarization of 0. We determine an upper limit of 3.7\,$\times$\,10$^{9}$\,cm$^{-3}$ for the electron density in the field lines. However, if a non-zero polarization fraction is eventually detected, suggesting that Faraday rotation in particular is ineffective, then the upper limit for the emitting region must be revised to $\leq$\,10$^{5}$--10$^{6}$\,cm$^{-3}$. Otherwise, another hinderence to the detection of polarization is the preferred orientation and topologies for the magnetic axes in order for reconnection to occur: either the dipoles are misaligned at a significant angle or else complex, higher-order magnetic structures are present at large stellar heights. In both cases, and for low-resolution observations specifically, reversed polarity footprints on each stellar surface will affect the measured polarization fraction. All things considered, the upper limit for the polarization in no way contradicts a synchrotron emission process.

The results of the simultaneous millimeter and optical monitoring reveal a particularly striking coincidence between the timing and duration of the multi-wavelength activity, but ultimately unravels little of the relationship between the dynamical and magnetospheric processes. We cannot distinguish between optical brightenings due to accretion events, reconnection processes, or a combination thereof. The current window for elevated millimeter activity is about 2.5 times smaller than the window for the optical brightenings. This preliminary statistic suggests that dynamically-induced accretion can occur independently; while each reconnection event thus far has been accompanied by an optical brightening that may, or may not, be associated with the accretion process only. Perhaps the only true test for the possible dependence of one process on the other, is to study similar binary systems where one effect is absent. We do suspect that this flaring phenomenon may be relevant to many similar T~Tauri binary systems of high eccentricity, as we expect to see when new millimeter instrumentation reduces the time required for large, multi-epoch surveys.


\begin{acknowledgements}

We would like to thank Vincent Pi\'etu and Jan-Martin Winters at IRAM for their assistance in scheduling and carrying out the observations, as well as Sascha Trippe for help with the PdBI data reduction and analysis. At CARMA we are grateful to Lee Mundy and Nikolaus Volgenau for their assistance in acquiring the data in Director's discretionary time. We would like to thank the observing staff at Teide Observatory for the IAC80 observations, including \'Alex Oscoz Abad, Cristina Zurita, and Rafael Barrena Delgado. At Wellesley College, we are grateful to Kim McLeod, Wendy Bauer, Steve Slivan and the undergraduate observers Kirsten Kelsey and Kathryn Neugent. Finally, we thank Y.~Boehler, A.~Dutrey, V.~Pi\'etu and S.~Guilloteau for communicating data prior to publication. Financial support for travel to Wellesley and IRAM for observational duties was provided by a Leids Sterrekunde Fonds grant. The research of DMS, AK, and MRH is supported through a VIDI grant from the Netherlands Organization for Scientific Research. 

\end{acknowledgements}

\bibliographystyle{aa}
\bibliography{15197}


\Online
\begin{appendix} 

\section{Optical data}

\begin{table*}[!th] 
\caption{Optical observing log} 
\label{optlog} 
\centering
\renewcommand{\footnoterule}{}  
\begin{minipage}[center]{16cm}
\begin{tabular}{c c c | c c c | c c c} 
\hline 
\hline
 JD+2450000 & Obs.$^{a}$ & V [mag]~ & ~JD+2450000 & Obs. & R [mag]~ & ~JD+2450000 & Obs. & I [mag] \\ 
\hline

 ~ & ~ & ~ & ~ & ~ & ~ & ~ & ~ & ~ \\
4803.640   &  W  &  13.64  &  4803.635  &  W  &  12.34  &  4803.661   &  W  &  11.12 \\
4809.513   &  W  &  13.70  &  4809.517  &  W  &  12.34  &  4809.519   &  W  &  11.07 \\
4809.534   &  W  &  13.59  &  4809.538  &  W  &  12.31  &  4809.540   &  W  &  11.07 \\
4813.484   &  W  &  13.59  &  4813.489  &  W  &  12.13  &  4813.491   &  W  &  11.03 \\
4813.506   &  W  &  13.77  &  4813.589  &  W  &  12.26  &  4813.591   &  W  &  11.06 \\
4813.585   &  W  &  13.49  &  4817.416  &  T  &  12.32  &  4817.418   &  T  &  11.10 \\
4817.414   &  T  &  13.66  &  4817.468  &  T  &  12.32  &  4817.469   &  T  &  11.11 \\
4817.466   &  T  &  13.67  &  4817.503  &  T  &  12.33  &  4817.504   &  T  &  11.10 \\
4817.501   &  T  &  13.67  &  4817.558  &  T  &  12.33  &  4817.560   &  T  &  11.20 \\
4817.557   &  T  &  13.68  &  4817.682  &  T  &  12.34  &  4817.683   &  T  &  11.11 \\
4817.680   &  T  &  13.67  &  4826.476  &  W  &  12.19  &  4826.472   &  W  &  10.99 \\
4826.480   &  W  &  13.37  &  4826.640  &  W  &  12.18  &  4826.473   &  W  &  11.02 \\
4826.632   &  W  &  13.51  &  4826.645  &  W  &  12.19  &  4826.651   &  W  &  11.06 \\
4826.812   &  W  &  13.57  &  4826.805  &  W  &  12.29  &  4826.801   &  W  &  11.12 \\
4828.451   &  T  &  13.53  &  4828.454  &  T  &  12.19  &  4828.455   &  T  &  11.00 \\
4828.522   &  T  &  13.52  &  4828.523  &  T  &  12.17  &  4828.524   &  T  &  11.00 \\
4828.591   &  T  &  13.49  &  4828.592  &  T  &  12.15  &  4828.593   &  T  &  10.98 \\
4829.414   &  T  &  13.47  &  4829.415  &  T  &  12.17  &  4829.430   &  T  &  10.99 \\
4829.427   &  T  &  13.49  &  4829.429  &  T  &  12.18  &  4829.459   &  W  &  11.05 \\
4829.464   &  W  &  13.54  &  4829.462  &  W  &  12.18  &  4829.470   &  T  &  11.01 \\
4829.468   &  T  &  13.53  &  4829.469  &  T  &  12.21  &  4829.471   &  W  &  11.02 \\
4829.502   &  W  &  13.51  &  4829.479  &  W  &  12.31  &  4829.505   &  W  &  11.06 \\
4829.506   &  T  &  13.55  &  4829.507  &  T  &  12.23  &  4829.508   &  T  &  11.02 \\
4829.513   &  W  &  13.66  &  4829.508  &  W  &  12.29  &  4829.547   &  T  &  11.03 \\
4829.521   &  W  &  13.64  &  4829.525  &  W  &  12.23  &  4829.548   &  W  &  11.06 \\
4829.543   &  W  &  13.69  &  4829.527  &  W  &  12.28  &  4829.560   &  W  &  11.09 \\
4829.544   &  T  &  13.57  &  4829.546  &  W  &  12.31  &  4829.591   &  W  &  11.09 \\
4829.555   &  W  &  13.54  &  4829.546  &  T  &  12.24  &  4829.596   &  T  &  11.00 \\
4829.585   &  W  &  13.67  &  4829.558  &  W  &  12.29  &  4829.604   &  W  &  11.08 \\
4829.593   &  T  &  13.52  &  4829.589  &  W  &  12.32  &  4829.613   &  W  &  11.09 \\
4829.599   &  W  &  13.65  &  4829.595  &  T  &  12.22  &  4829.635   &  W  &  11.08 \\
4829.607   &  W  &  13.68  &  4829.603  &  W  &  12.32  &  4829.667   &  W  &  11.07 \\
4829.629   &  W  &  13.69  &  4829.611  &  W  &  12.30  &  4829.707   &  W  &  11.06 \\
4829.661   &  W  &  13.67  &  4829.633  &  W  &  12.30  &  4829.728   &  W  &  11.11 \\
4829.702   &  W  &  13.79  &  4829.665  &  W  &  12.31  &  4829.756   &  W  &  11.11 \\
4829.722   &  W  &  13.68  &  4829.726  &  W  &  12.32  &  4829.794   &  W  &  11.11 \\
4829.798   &  W  &  13.69  &  4829.753  &  W  &  12.32  &  4831.440   &  W  &  11.10 \\
4831.448   &  W  &  13.77  &  4829.796  &  W  &  12.32  &  4831.463   &  W  &  11.08 \\
4831.468   &  W  &  13.75  &  4831.444  &  W  &  12.31  &  4831.479   &  W  &  11.11 \\
4831.485   &  W  &  13.67  &  4831.465  &  W  &  12.33  &  4831.488   &  W  &  11.11 \\
4831.493   &  W  &  13.70  &  4831.481  &  W  &  12.33  &  4831.502   &  W  &  11.10 \\
4831.508   &  W  &  13.68  &  4831.489  &  W  &  12.31  &  4834.461   &  T  &  11.07 \\
4834.454   &  T  &  13.62  &  4831.504  &  W  &  12.32  &  4834.536   &  T  &  11.08 \\
4834.529   &  T  &  13.62  &  4834.458  &  T  &  12.29  &  4834.603   &  T  &  11.07 \\
4834.596   &  T  &  13.67  &  4834.533  &  T  &  12.29  &  4834.661   &  T  &  11.08 \\
4834.654   &  T  &  13.61  &  4834.600  &  T  &  12.30  &  4836.390   &  T  &  11.10 \\
4836.396   &  T  &  13.69  &  4834.658  &  T  &  12.30  &  4836.398   &  T  &  11.10 \\
4836.404   &  T  &  13.68  &  4836.393  &  T  &  12.34  &  4836.407   &  T  &  11.10 \\
4836.413   &  T  &  13.69  &  4836.401  &  T  &  12.34  &  4836.519   &  T  &  11.09 \\
4836.525   &  T  &  13.67  &  4836.410  &  T  &  12.34  &  4836.528   &  T  &  11.09 \\
4836.534   &  T  &  13.66  &  4836.522  &  T  &  12.33  &  4836.536   &  T  &  11.09 \\
4836.542   &  T  &  13.67  &  4836.530  &  T  &  12.33  &  4842.375   &  T  &  11.00 \\
4842.369   &  T  &  13.49  &  4836.539  &  T  &  12.32  &  4842.387   &  T  &  11.00 \\
4842.380   &  T  &  13.46  &  4842.372  &  T  &  12.17  &  4842.517   &  T  &  10.96 \\
4842.510   &  T  &  13.34  &  4842.384  &  T  &  12.15  &  4842.527   &  T  &  10.96 \\
4842.520   &  T  &  13.34  &  4842.513  &  T  &  12.08  &  5208.351   &  T  &  10.91 \\
5208.360   &  T  &  13.30  &  4842.523	&  T  &  12.08  &  5208.474   &  T  &  10.91 \\
5208.482   &  T  &  13.31  &  5208.357  &  T  &  12.04  &  5208.592   &  T  &  10.90 \\ 
5208.597   &  T  &  13.31  &  5208.480  &  T  &  12.05  &  5210.532   &  T  &  11.06 \\ 
5210.540   &  T  &  13.61  &  5208.596  &  T  &  12.07  &  5210.542   &  T  &  11.06 \\ 
5210.550   &  T  &  13.61  &  5210.537  &  T  &  12.32  &  5210.553   &  T  &  11.06 \\ 
5210.560   &  T  &  13.62  &  5210.547  &  T  &  12.31  &  5210.563   &  T  &  11.06 \\
5210.571   &  T  &  13.62  &  5210.558  &  T  &  12.31  &  ~          &  ~  &  ~     \\ 
~          & ~   &  ~      &  5210.568  &  T  &  12.31  &  ~          &  ~  &  ~     \\ 
~ & ~ & ~ & ~ & ~ & ~ & ~ & ~ & ~ \\
\hline
\multicolumn{9}{l}{~} \\
\multicolumn{9}{l}{$^{a}$ To indicate the Teide Observatory IAC-80 telescope in the Canary Islands (Spain) we use a `T', and for the}\\
\multicolumn{9}{l}{Wellesley 0.6-meter telescope at Wellesley College in Wellesley, Massachusetts (USA) we use a `W'.}\\

\end{tabular} 
\end{minipage}
\end{table*} 


\begin{table}[!ht] 
\begin{minipage}{\columnwidth}
\caption{Comparison stars}
\label{comp} 
\centering 
\renewcommand{\footnoterule}{}  
\begin{tabular}{c c c c} 
\hline 
\hline
Number\footnote{The numbers correspond to the labels in Fig.~\ref{compfig}} & V [mag] & R [mag] & I [mag] \\
\hline
1    & 15.79 & 14.68 & 13.81 \\
2    & 14.86 & 13.95 & 13.18 \\
3    & 14.43 & 13.33 & 12.27 \\
4    & 13.72 & 12.61 & 11.66 \\
5    & 11.84 & 11.05 & 10.43 \\
6    & 15.30 & 14.12 & 13.19 \\
\hline 
\end{tabular} 
\end{minipage} 
\end{table} 


\begin{figure}[hbt!]
\centering
\includegraphics[width=\columnwidth]{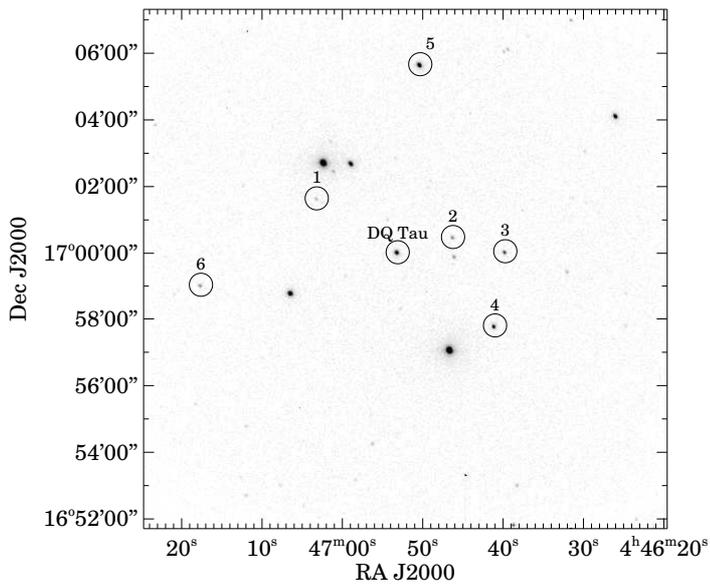}
\caption{The comparison stars used to derive the differential photometry for DQ Tau. The figure shows the complete 15.6$'\,{\times}\,$15.6$'$ field of view for the Wellesley 0.6-meter telescope. This I-band image was obtained on 2008 December 25.}
\label{compfig}
\end{figure}


\clearpage
\newpage

\section{Millimeter data}

\begin{table*}[h!]
\caption{Julian day, flux, and uncertainty values for the millimeter light curve on 2008 December 28-29, recorded at a central frequency of 90.0\,GHz with the IRAM PdBI. \label{dec08iram} } 
\centering 
\begin{tabular}{ccc|ccc|ccc|ccc}
\hline 
\hline
 ~Julian Day & Flux & $\sigma$ &  ~Julian Day & Flux & $\sigma$ &  ~Julian Day & Flux & $\sigma$ &  ~Julian Day & Flux & $\sigma$ \\
 ($+$2450000) & [mJy] & [mJy] & ($+$2450000) & [mJy] & [mJy] & ($+$2450000) & [mJy] & [mJy] &  ($+$2450000) & [mJy] & [mJy] \\
\hline
\noalign{\smallskip}

4829.290  &  44.4  &  1.0  &  4829.423  &  26.8  &  1.0  &  4829.538  &  25.5  &  1.0 &  ~ & ~ & ~ \\
4829.291  &  42.9  &  1.0  &  4829.424  &  28.9  &  1.0  &  4829.539  &  25.9  &  1.0 &  ~ & ~ & ~ \\
4829.293  &  42.8  &  1.0  &  4829.425  &  28.0  &  1.0  &  4829.540  &  22.8  &  1.1 &  ~ & ~ & ~ \\
4829.294  &  42.5  &  1.0  &  4829.438  &  30.4  &  1.0  &  4829.541  &  25.7  &  1.1 &  ~ & ~ & ~ \\
4829.295  &  42.5  &  1.0  &  4829.439  &  29.1  &  1.0  &  4829.542  &  24.2  &  1.1 &  ~ & ~ & ~ \\
4829.296  &  42.5  &  1.0  &  4829.440  &  27.5  &  1.0  &  4829.543  &  24.8  &  1.1 &  ~ & ~ & ~ \\
4829.297  &  43.1  &  1.0  &  4829.441  &  30.5  &  1.0  &  4829.554  &  22.5  &  1.1 &  ~ & ~ & ~ \\
4829.298  &  39.6  &  1.0  &  4829.443  &  27.0  &  1.0  &  4829.555  &  23.5  &  1.1 &  ~ & ~ & ~ \\
4829.299  &  39.9  &  1.0  &  4829.444  &  30.2  &  1.0  &  4829.557  &  23.3  &  1.1 &  ~ & ~ & ~ \\
4829.300  &  40.5  &  1.0  &  4829.445  &  26.6  &  1.0  &  4829.558  &  23.5  &  1.1 &  ~ & ~ & ~ \\
4829.301  &  40.3  &  1.0  &  4829.446  &  28.3  &  1.0  &  4829.559  &  25.5  &  1.1 &  ~ & ~ & ~ \\
4829.302  &  38.9  &  1.0  &  4829.447  &  28.7  &  1.0  &  4829.560  &  24.5  &  1.1 &  ~ & ~ & ~ \\
4829.303  &  36.9  &  1.0  &  4829.448  &  28.2  &  1.0  &  4829.561  &  23.1  &  1.1 &  ~ & ~ & ~ \\
4829.304  &  39.9  &  1.0  &  4829.449  &  29.7  &  1.0  &  4829.562  &  22.6  &  1.1 &  ~ & ~ & ~ \\
4829.305  &  37.4  &  1.0  &  4829.450  &  28.7  &  1.0  &  4829.563  &  23.7  &  1.1 &  ~ & ~ & ~ \\
4829.315  &  37.0  &  1.0  &  4829.451  &  27.5  &  1.0  &  4829.564  &  21.6  &  1.1 &  ~ & ~ & ~ \\
4829.316  &  36.9  &  1.0  &  4829.452  &  31.4  &  1.0  &  4829.565  &  23.3  &  1.1 &  ~ & ~ & ~ \\
4829.317  &  37.6  &  1.0  &  4829.453  &  25.5  &  1.0  &  4829.566  &  22.7  &  1.1 &  ~ & ~ & ~ \\
4829.318  &  36.3  &  1.0  &  4829.461  &  29.7  &  1.0  &  4829.567  &  22.2  &  1.1 &  ~ & ~ & ~ \\
4829.319  &  38.1  &  1.0  &  4829.462  &  27.9  &  1.0  &  4829.568  &  21.4  &  1.1 &  ~ & ~ & ~ \\
4829.320  &  37.6  &  1.0  &  4829.463  &  27.5  &  1.0  &  4829.569  &  22.8  &  1.1 &  ~ & ~ & ~ \\
4829.321  &  38.0  &  1.0  &  4829.464  &  30.6  &  1.0  &  4829.581  &  22.4  &  2.1 &  ~ & ~ & ~ \\
4829.322  &  36.9  &  1.0  &  4829.465  &  29.2  &  1.0  &  4829.582  &  23.0  &  2.1 &  ~ & ~ & ~ \\
4829.323  &  36.8  &  1.0  &  4829.466  &  25.3  &  1.0  &  4829.583  &  24.4  &  2.1 &  ~ & ~ & ~ \\
4829.324  &  37.6  &  1.0  &  4829.467  &  30.5  &  1.0  &  4829.584  &  20.3  &  2.1 &  ~ & ~ & ~ \\
4829.325  &  37.8  &  1.0  &  4829.468  &  27.0  &  1.0  &  4829.585  &  25.2  &  2.2 &  ~ & ~ & ~ \\
4829.326  &  37.2  &  1.0  &  4829.469  &  30.3  &  1.0  &  4829.586  &  24.1  &  2.2 &  ~ & ~ & ~ \\
4829.327  &  34.4  &  1.0  &  4829.470  &  28.0  &  1.0  &  4829.616  &  13.6  &  2.4 &  ~ & ~ & ~ \\
4829.328  &  34.0  &  1.0  &  4829.471  &  30.5  &  1.0  &  4829.617  &  20.7  &  2.5 &  ~ & ~ & ~ \\
4829.329  &  35.5  &  1.0  &  4829.472  &  28.9  &  1.0  &  4829.618  &  20.9  &  2.5 &  ~ & ~ & ~ \\
4829.386  &  31.2  &  1.0  &  4829.473  &  27.1  &  1.0  &  4829.619  &  21.0  &  2.5 &  ~ & ~ & ~ \\
4829.387  &  32.5  &  1.0  &  4829.474  &  28.2  &  1.0  &  4829.620  &  21.6  &  2.5 &  ~ & ~ & ~ \\
4829.388  &  29.6  &  1.0  &  4829.475  &  28.8  &  1.0  &  4829.622  &  19.5  &  2.5 &  ~ & ~ & ~ \\
4829.389  &  28.6  &  1.0  &  4829.488  &  28.1  &  1.0  &  4829.623  &  21.4  &  2.6 &  ~ & ~ & ~ \\
4829.391  &  31.6  &  1.0  &  4829.489  &  28.5  &  1.0  &  4829.624  &  24.2  &  2.6 &  ~ & ~ & ~ \\
4829.392  &  31.3  &  1.0  &  4829.490  &  28.9  &  1.0  &  4829.625  &  13.6  &  2.6 &  ~ & ~ & ~ \\
4829.393  &  31.0  &  1.0  &  4829.491  &  28.3  &  1.0  &  4829.626  &  18.8  &  2.6 &  ~ & ~ & ~ \\
4829.394  &  32.3  &  1.0  &  4829.492  &  28.0  &  1.0  &  4829.627  &  19.7  &  2.6 &  ~ & ~ & ~ \\
4829.395  &  31.5  &  1.0  &  4829.493  &  26.0  &  1.0  &  4829.628  &  22.6  &  2.7 &  ~ & ~ & ~ \\
4829.396  &  31.1  &  1.0  &  4829.494  &  25.0  &  1.0  &  4829.629  &  19.3  &  2.7 &  ~ & ~ & ~ \\
4829.397  &  32.0  &  1.0  &  4829.495  &  26.7  &  1.0  &  4829.630  &  21.7  &  2.7 &  ~ & ~ & ~ \\
4829.398  &  29.8  &  1.0  &  4829.496  &  27.6  &  1.0  &  4829.631  &  17.6  &  2.7 &  ~ & ~ & ~ \\
4829.399  &  33.1  &  1.0  &  4829.497  &  27.6  &  1.0  &  ~ & ~ & ~ &  ~ & ~ & ~ \\
4829.400  &  32.0  &  1.0  &  4829.498  &  27.9  &  1.0  &  ~ & ~ & ~ &  ~ & ~ & ~ \\
4829.401  &  34.4  &  1.0  &  4829.499  &  25.0  &  1.0  &  ~ & ~ & ~ &  ~ & ~ & ~ \\
4829.410  &  32.1  &  1.0  &  4829.500  &  24.3  &  1.0  &  ~ & ~ & ~ &  ~ & ~ & ~ \\
4829.411  &  30.2  &  1.0  &  4829.501  &  26.6  &  1.0  &  ~ & ~ & ~ &  ~ & ~ & ~ \\
4829.412  &  29.7  &  1.0  &  4829.502  &  26.2  &  1.0  &  ~ & ~ & ~ &  ~ & ~ & ~ \\
4829.413  &  29.3  &  1.0  &  4829.529  &  22.4  &  1.0  &  ~ & ~ & ~ &  ~ & ~ & ~ \\
4829.414  &  30.7  &  1.0  &  4829.530  &  24.4  &  1.0  &  ~ & ~ & ~ &  ~ & ~ & ~ \\
4829.415  &  30.4  &  1.0  &  4829.531  &  24.7  &  1.0  &  ~ & ~ & ~ &  ~ & ~ & ~ \\
4829.416  &  30.7  &  1.0  &  4829.532  &  23.5  &  1.0  &  ~ & ~ & ~ &  ~ & ~ & ~ \\
4829.417  &  30.2  &  1.0  &  4829.533  &  25.8  &  1.0  &  ~ & ~ & ~ &  ~ & ~ & ~ \\
4829.418  &  29.2  &  1.0  &  4829.534  &  24.2  &  1.0  &  ~ & ~ & ~ &  ~ & ~ & ~ \\
4829.420  &  30.2  &  1.0  &  4829.535  &  26.4  &  1.0  &  ~ & ~ & ~ &  ~ & ~ & ~ \\
4829.421  &  27.9  &  1.0  &  4829.536  &  24.9  &  1.0  &  ~ & ~ & ~ &  ~ & ~ & ~ \\
4829.422  &  28.6  &  1.0  &  4829.537  &  22.5  &  1.0  &  ~ & ~ & ~ &  ~ & ~ & ~ \\

\hline
\end{tabular}
\end{table*}

\begin{table*}
\caption{Julian day, flux, and uncertainty values for the millimeter light curve on 2009 March 17, recorded at a central frequency of 90.0\,GHz with the IRAM PdBI. \label{mar09iram} } 
\centering 
\begin{tabular}{ccc|ccc|ccc|ccc}
\hline 
\hline
 ~Julian Day & Flux & $\sigma$ &  ~Julian Day & Flux & $\sigma$ &  ~Julian Day & Flux & $\sigma$ &  ~Julian Day & Flux & $\sigma$ \\
 ($+$2450000) & [mJy] & [mJy] & ($+$2450000) & [mJy] & [mJy] & ($+$2450000) & [mJy] & [mJy] &  ($+$2450000) & [mJy] & [mJy] \\
\hline
\noalign{\smallskip}

4908.028  &  12.8  &  1.7  &  4908.110  &  15.4  &  1.3  &  4908.199  &  12.5  &  0.9  &  4908.288  &  10.5  &  0.9  \\
4908.029  &  9.4   &  1.7  &  4908.111  &  12.3  &  1.2  &  4908.200  &  9.4   &  0.9  &  4908.289  &  12.4  &  0.9  \\
4908.030  &  17.0  &  1.7  &  4908.112  &  10.7  &  1.2  &  4908.201  &  12.4  &  0.9  &  4908.290  &  8.5   &  0.9  \\
4908.031  &  12.1  &  1.7  &  4908.122  &  12.1  &  1.0  &  4908.202  &  11.6  &  0.9  &  4908.291  &  10.5  &  0.9  \\
4908.032  &  11.6  &  1.7  &  4908.123  &  13.3  &  1.0  &  4908.203  &  11.1  &  0.9  &  4908.292  &  11.4  &  0.9  \\
4908.033  &  11.8  &  1.7  &  4908.124  &  13.5  &  1.0  &  4908.204  &  12.2  &  0.9  &  4908.293  &  11.6  &  0.9  \\
4908.034  &  8.5   &  1.6  &  4908.125  &  13.1  &  1.0  &  4908.214  &  11.8  &  0.8  &  4908.294  &  13.0  &  0.9  \\
4908.035  &  9.4   &  1.6  &  4908.126  &  13.9  &  1.0  &  4908.215  &  11.4  &  0.8  &  4908.295  &  12.6  &  0.9  \\
4908.036  &  10.1  &  1.6  &  4908.127  &  12.1  &  1.0  &  4908.216  &  12.6  &  0.8  &  4908.296  &  10.4  &  0.9  \\
4908.037  &  12.9  &  1.6  &  4908.128  &  14.0  &  1.0  &  4908.217  &  13.5  &  0.8  &  4908.305  &  11.2  &  0.9  \\
4908.038  &  15.4  &  1.6  &  4908.129  &  9.4   &  1.0  &  4908.218  &  11.9  &  0.9  &  4908.306  &  12.1  &  0.9  \\
4908.039  &  13.0  &  1.6  &  4908.130  &  13.7  &  1.0  &  4908.219  &  12.1  &  0.9  &  4908.307  &  11.5  &  0.9  \\
4908.040  &  11.5  &  1.6  &  4908.131  &  11.6  &  1.0  &  4908.220  &  10.6  &  0.9  &  4908.308  &  11.4  &  0.9  \\
4908.041  &  13.9  &  1.6  &  4908.132  &  12.4  &  1.0  &  4908.221  &  12.6  &  0.9  &  4908.309  &  12.0  &  0.9  \\
4908.042  &  15.6  &  1.6  &  4908.133  &  12.7  &  1.0  &  4908.222  &  12.4  &  0.9  &  4908.311  &  13.3  &  0.9  \\
4908.050  &  13.2  &  1.6  &  4908.134  &  13.4  &  1.0  &  4908.223  &  9.4   &  0.9  &  4908.312  &  9.9   &  0.9  \\
4908.051  &  13.1  &  1.6  &  4908.135  &  10.9  &  1.0  &  4908.224  &  12.0  &  0.9  &  4908.313  &  12.3  &  0.9  \\
4908.052  &  13.5  &  1.6  &  4908.136  &  12.5  &  1.0  &  4908.225  &  12.8  &  0.9  &  4908.314  &  13.0  &  0.9  \\
4908.053  &  13.9  &  1.6  &  4908.144  &  11.9  &  0.9  &  4908.226  &  10.3  &  0.9  &  4908.315  &  9.3   &  1.0  \\
4908.054  &  13.9  &  1.6  &  4908.145  &  10.9  &  0.9  &  4908.227  &  13.2  &  0.9  &  4908.316  &  13.1  &  1.0  \\
4908.055  &  11.0  &  1.6  &  4908.146  &  11.2  &  0.9  &  4908.228  &  9.8   &  0.9  &  4908.317  &  12.5  &  1.0  \\
4908.056  &  12.0  &  1.6  &  4908.147  &  11.8  &  0.9  &  4908.235  &  13.2  &  0.8  &  4908.318  &  10.8  &  1.0  \\
4908.057  &  9.0   &  1.6  &  4908.148  &  10.1  &  0.9  &  4908.237  &  11.5  &  0.8  &  4908.319  &  11.2  &  1.0  \\
4908.058  &  10.6  &  1.6  &  4908.149  &  14.1  &  0.9  &  4908.238  &  10.8  &  0.8  &  4908.320  &  10.0  &  1.0  \\
4908.059  &  13.0  &  1.6  &  4908.150  &  12.1  &  0.9  &  4908.239  &  9.8   &  0.8  &  4908.327  &  11.1  &  1.0  \\
4908.060  &  12.0  &  1.6  &  4908.151  &  11.1  &  0.9  &  4908.240  &  10.9  &  0.8  &  4908.328  &  12.7  &  1.0  \\
4908.061  &  10.6  &  1.6  &  4908.152  &  12.1  &  0.9  &  4908.241  &  12.1  &  0.8  &  4908.329  &  12.3  &  1.0  \\
4908.062  &  11.0  &  1.6  &  4908.153  &  11.0  &  0.9  &  4908.242  &  9.5   &  0.8  &  4908.330  &  11.4  &  1.0  \\
4908.063  &  10.4  &  1.6  &  4908.154  &  11.6  &  0.9  &  4908.243  &  11.0  &  0.8  &  4908.331  &  10.5  &  1.0  \\
4908.064  &  11.9  &  1.6  &  4908.155  &  12.3  &  0.9  &  4908.244  &  11.6  &  0.8  &  4908.332  &  11.9  &  1.0  \\
4908.076  &  14.1  &  1.1  &  4908.156  &  13.6  &  0.9  &  4908.245  &  11.3  &  0.8  &  4908.333  &  11.0  &  1.0  \\
4908.077  &  12.5  &  1.1  &  4908.157  &  10.3  &  0.9  &  4908.246  &  11.3  &  0.8  &  4908.334  &  12.6  &  1.0  \\
4908.078  &  11.0  &  1.1  &  4908.158  &  11.2  &  0.9  &  4908.247  &  10.9  &  0.8  &  4908.335  &  12.3  &  1.0  \\
4908.079  &  13.0  &  1.1  &  4908.168  &  12.0  &  0.9  &  4908.248  &  12.5  &  0.8  &  4908.336  &  12.9  &  1.0  \\
4908.080  &  11.3  &  1.1  &  4908.169  &  12.1  &  0.9  &  4908.249  &  12.8  &  0.8  &  4908.337  &  12.1  &  1.0  \\
4908.081  &  12.2  &  1.1  &  4908.170  &  11.9  &  0.9  &  4908.250  &  10.9  &  0.8  &  4908.338  &  13.6  &  1.0  \\
4908.082  &  14.8  &  1.1  &  4908.171  &  13.3  &  0.9  &  4908.260  &  11.3  &  0.9  &  4908.340  &  12.8  &  1.0  \\
4908.083  &  11.4  &  1.1  &  4908.172  &  11.8  &  0.9  &  4908.261  &  12.5  &  0.9  &  4908.341  &  16.1  &  1.0  \\
4908.084  &  13.7  &  1.1  &  4908.173  &  11.4  &  0.9  &  4908.262  &  10.5  &  0.9  &  4908.342  &  11.6  &  1.0  \\
4908.085  &  11.4  &  1.1  &  4908.174  &  11.5  &  0.9  &  4908.263  &  12.8  &  0.9  &  4908.351  &  19.2  &  1.1  \\
4908.086  &  12.8  &  1.1  &  4908.175  &  12.6  &  0.9  &  4908.264  &  12.8  &  0.9  &  4908.352  &  18.6  &  1.1  \\
4908.087  &  13.8  &  1.1  &  4908.176  &  12.1  &  0.9  &  4908.265  &  13.9  &  0.9  &  4908.353  &  18.0  &  1.1  \\
4908.088  &  14.5  &  1.1  &  4908.177  &  12.9  &  0.9  &  4908.266  &  12.5  &  0.9  &  4908.354  &  17.2  &  1.1  \\
4908.089  &  13.7  &  1.1  &  4908.178  &  11.8  &  0.9  &  4908.267  &  11.1  &  0.9  &  4908.355  &  20.9  &  1.1  \\
4908.091  &  13.5  &  1.1  &  4908.179  &  10.7  &  0.9  &  4908.268  &  10.3  &  0.9  &  4908.356  &  16.4  &  1.1  \\
4908.098  &  8.1   &  1.3  &  4908.180  &  13.0  &  0.9  &  4908.269  &  11.8  &  0.9  &  4908.357  &  18.7  &  1.1  \\
4908.099  &  12.4  &  1.3  &  4908.181  &  12.0  &  0.9  &  4908.270  &  9.8   &  0.9  &  4908.358  &  16.6  &  1.1  \\
4908.100  &  12.2  &  1.3  &  4908.182  &  12.3  &  0.9  &  4908.271  &  12.3  &  0.9  &  4908.359  &  16.4  &  1.1  \\
4908.101  &  13.0  &  1.3  &  4908.190  &  11.8  &  0.9  &  4908.272  &  12.3  &  0.9  &  4908.360  &  18.1  &  1.1  \\
4908.102  &  7.9   &  1.3  &  4908.191  &  11.8  &  0.9  &  4908.273  &  15.2  &  0.9  &  4908.361  &  17.9  &  1.1  \\
4908.103  &  15.9  &  1.3  &  4908.192  &  13.8  &  0.9  &  4908.274  &  12.2  &  0.9  &  4908.363  &  15.6  &  1.4  \\
4908.104  &  12.9  &  1.3  &  4908.193  &  11.9  &  0.9  &  4908.281  &  11.2  &  0.9  &  4908.364  &  17.9  &  1.4  \\
4908.105  &  10.4  &  1.3  &  4908.194  &  12.9  &  0.9  &  4908.282  &  10.0  &  0.9  &  4908.365  &  16.9  &  1.4  \\
4908.106  &  11.7  &  1.3  &  4908.195  &  10.4  &  0.9  &  4908.283  &  10.0  &  0.9  &  4908.366  &  16.0  &  1.4  \\
4908.107  &  12.2  &  1.3  &  4908.196  &  10.7  &  0.9  &  4908.284  &  13.4  &  0.9  &   ~ &  ~ &  ~ \\
4908.108  &  11.5  &  1.3  &  4908.197  &  12.1  &  0.9  &  4908.285  &  13.0  &  0.9  &   ~ &  ~ &  ~ \\
4908.109  &  11.2  &  1.3  &  4908.198  &  12.4  &  0.9  &  4908.286  &  9.9   &  0.9  &   ~ &  ~ &  ~ \\

\hline
\end{tabular}
\end{table*}

\begin{table*}
\caption{Julian day, flux, and uncertainty values for the millimeter light curve on 2010 January 11-12, recorded at a central frequency of 90.0\,GHz with the IRAM PdBI. \label{jan10iram} } 
\centering 
\begin{tabular}{ccc|ccc|ccc|ccc}
\hline 
\hline
 ~Julian Day & Flux & $\sigma$ &  ~Julian Day & Flux & $\sigma$ &  ~Julian Day & Flux & $\sigma$ &  ~Julian Day & Flux & $\sigma$ \\
 ($+$2450000) & [mJy] & [mJy] & ($+$2450000) & [mJy] & [mJy] & ($+$2450000) & [mJy] & [mJy] &  ($+$2450000) & [mJy] & [mJy] \\
\hline
\noalign{\smallskip}

 5208.266 &  62.6  &  1.0  &  5208.345  &  37.4  &  1.0  &  5208.433  &  49.6  &  1.3  &  5208.588  &  30.5  &  1.5  \\
 5208.267 &  64.6  &  1.0  &  5208.346  &  38.9  &  1.0  &  5208.434  &  50.2  &  1.3  &  5208.589  &  27.9  &  1.5  \\
 5208.268 &  62.4  &  1.0  &  5208.347  &  40.5  &  1.0  &  5208.435  &  51.1  &  1.3  &  5208.590  &  31.4  &  1.5  \\
 5208.269 &  62.2  &  1.0  &  5208.357  &  34.6  &  1.0  &  5208.436  &  52.9  &  1.3  &  5208.591  &  29.2  &  1.5  \\
 5208.270 &  60.3  &  1.0  &  5208.358  &  39.2  &  1.0  &  5208.438  &  50.7  &  1.3  &  5208.592  &  31.6  &  1.5  \\
 5208.271 &  60.3  &  1.0  &  5208.359  &  37.7  &  1.0  &  5208.439  &  50.8  &  1.3  &  5208.593  &  25.7  &  1.6  \\
 5208.272 &  60.3  &  1.0  &  5208.360  &  38.0  &  1.0  &  5208.515  &  33.0  &  1.1  &  5208.595  &  30.6  &  1.6  \\
 5208.273 &  58.8  &  1.0  &  5208.361  &  41.1  &  1.0  &  5208.516  &  35.8  &  1.1  &  5208.596  &  27.6  &  1.6  \\
 5208.274 &  60.3  &  1.0  &  5208.362  &  38.4  &  1.0  &  5208.517  &  34.0  &  1.1  &  5208.597  &  30.7  &  1.6  \\
 5208.275 &  58.8  &  1.0  &  5208.363  &  37.7  &  1.0  &  5208.518  &  32.0  &  1.1  &   ~ &  ~ &  ~ \\
 5208.276 &  59.9  &  1.0  &  5208.364  &  36.5  &  1.0  &  5208.519  &  33.1  &  1.1  &   ~ &  ~ &  ~ \\
 5208.277 &  59.9  &  1.0  &  5208.365  &  39.9  &  1.0  &  5208.520  &  34.5  &  1.1  &   ~ &  ~ &  ~ \\
 5208.278 &  60.2  &  1.0  &  5208.366  &  36.8  &  1.0  &  5208.521  &  32.9  &  1.1  &   ~ &  ~ &  ~ \\
 5208.279 &  61.8  &  1.0  &  5208.367  &  38.8  &  1.0  &  5208.522  &  31.5  &  1.1  &   ~ &  ~ &  ~ \\
 5208.280 &  59.7  &  1.0  &  5208.368  &  35.7  &  1.0  &  5208.523  &  31.6  &  1.1  &   ~ &  ~ &  ~ \\
 5208.287 &  58.9  &  1.0  &  5208.369  &  37.3  &  1.0  &  5208.524  &  33.9  &  1.1  &   ~ &  ~ &  ~ \\
 5208.288 &  56.6  &  1.0  &  5208.370  &  37.5  &  1.0  &  5208.526  &  31.9  &  1.1  &   ~ &  ~ &  ~ \\
 5208.289 &  57.4  &  1.0  &  5208.371  &  36.1  &  1.0  &  5208.527  &  30.9  &  1.1  &   ~ &  ~ &  ~ \\
 5208.290 &  58.7  &  1.0  &  5208.378  &  33.4  &  1.0  &  5208.528  &  33.5  &  1.2  &   ~ &  ~ &  ~ \\
 5208.291 &  57.0  &  1.0  &  5208.379  &  35.4  &  1.0  &  5208.529  &  31.2  &  1.2  &   ~ &  ~ &  ~ \\
 5208.292 &  58.3  &  1.0  &  5208.380  &  37.3  &  1.0  &  5208.530  &  29.7  &  1.2  &   ~ &  ~ &  ~ \\
 5208.294 &  54.8  &  1.0  &  5208.382  &  36.2  &  1.0  &  5208.537  &  30.9  &  1.2  &   ~ &  ~ &  ~ \\
 5208.295 &  55.3  &  1.0  &  5208.383  &  39.1  &  1.0  &  5208.538  &  31.5  &  1.2  &   ~ &  ~ &  ~ \\
 5208.296 &  55.3  &  1.0  &  5208.384  &  37.3  &  1.0  &  5208.539  &  30.7  &  1.2  &   ~ &  ~ &  ~ \\
 5208.297 &  54.4  &  1.0  &  5208.385  &  36.2  &  1.0  &  5208.540  &  32.0  &  1.2  &   ~ &  ~ &  ~ \\
 5208.298 &  54.0  &  1.0  &  5208.386  &  40.5  &  1.0  &  5208.541  &  31.3  &  1.2  &   ~ &  ~ &  ~ \\
 5208.299 &  51.1  &  1.0  &  5208.387  &  37.0  &  1.0  &  5208.542  &  30.5  &  1.2  &   ~ &  ~ &  ~ \\
 5208.300 &  52.8  &  1.0  &  5208.388  &  40.2  &  1.0  &  5208.543  &  33.5  &  1.2  &   ~ &  ~ &  ~ \\
 5208.301 &  51.4  &  1.0  &  5208.389  &  39.7  &  1.0  &  5208.544  &  28.5  &  1.2  &   ~ &  ~ &  ~ \\
 5208.302 &  51.8  &  1.0  &  5208.390  &  37.5  &  1.0  &  5208.545  &  27.9  &  1.2  &   ~ &  ~ &  ~ \\
 5208.311 &  50.0  &  1.0  &  5208.391  &  37.6  &  1.0  &  5208.546  &  30.6  &  1.2  &   ~ &  ~ &  ~ \\
 5208.312 &  48.1  &  1.0  &  5208.392  &  38.9  &  1.0  &  5208.547  &  28.2  &  1.2  &   ~ &  ~ &  ~ \\
 5208.313 &  47.7  &  1.0  &  5208.393  &  37.4  &  1.0  &  5208.548  &  30.5  &  1.2  &   ~ &  ~ &  ~ \\
 5208.314 &  45.5  &  1.0  &  5208.402  &  38.3  &  1.0  &  5208.549  &  28.2  &  1.2  &   ~ &  ~ &  ~ \\
 5208.315 &  46.8  &  1.0  &  5208.403  &  37.3  &  1.0  &  5208.550  &  30.5  &  1.2  &   ~ &  ~ &  ~ \\
 5208.316 &  46.9  &  1.0  &  5208.404  &  39.1  &  1.0  &  5208.551  &  26.9  &  1.2  &   ~ &  ~ &  ~ \\
 5208.317 &  46.1  &  1.0  &  5208.405  &  36.6  &  1.0  &  5208.560  &  28.2  &  1.3  &   ~ &  ~ &  ~ \\
 5208.318 &  45.5  &  1.0  &  5208.406  &  40.3  &  1.0  &  5208.561  &  27.6  &  1.3  &   ~ &  ~ &  ~ \\
 5208.320 &  44.7  &  1.0  &  5208.408  &  40.6  &  1.0  &  5208.563  &  30.0  &  1.3  &   ~ &  ~ &  ~ \\
 5208.321 &  44.8  &  1.0  &  5208.409  &  39.1  &  1.0  &  5208.564  &  30.1  &  1.4  &   ~ &  ~ &  ~ \\
 5208.322 &  42.9  &  1.0  &  5208.410  &  37.2  &  1.0  &  5208.565  &  27.1  &  1.4  &   ~ &  ~ &  ~ \\
 5208.323 &  47.8  &  1.0  &  5208.411  &  39.4  &  1.0  &  5208.566  &  25.8  &  1.4  &   ~ &  ~ &  ~ \\
 5208.324 &  45.7  &  1.0  &  5208.412  &  38.9  &  1.0  &  5208.567  &  26.4  &  1.4  &   ~ &  ~ &  ~ \\
 5208.325 &  43.8  &  1.0  &  5208.413  &  38.2  &  1.0  &  5208.568  &  28.7  &  1.4  &   ~ &  ~ &  ~ \\
 5208.326 &  44.4  &  1.0  &  5208.414  &  41.1  &  1.0  &  5208.569  &  31.4  &  1.4  &   ~ &  ~ &  ~ \\
 5208.333 &  42.6  &  1.0  &  5208.415  &  38.6  &  1.0  &  5208.570  &  29.0  &  1.4  &   ~ &  ~ &  ~ \\
 5208.334 &  41.1  &  1.0  &  5208.416  &  42.7  &  1.0  &  5208.571  &  29.2  &  1.4  &   ~ &  ~ &  ~ \\
 5208.335 &  40.3  &  1.0  &  5208.417  &  41.4  &  1.0  &  5208.572  &  29.1  &  1.4  &   ~ &  ~ &  ~ \\
 5208.336 &  40.5  &  1.0  &  5208.424  &  49.5  &  1.3  &  5208.573  &  25.4  &  1.4  &   ~ &  ~ &  ~ \\
 5208.337 &  43.8  &  1.0  &  5208.425  &  43.5  &  1.3  &  5208.574  &  27.4  &  1.4  &   ~ &  ~ &  ~ \\
 5208.338 &  43.8  &  1.0  &  5208.426  &  47.5  &  1.3  &  5208.575  &  28.2  &  1.4  &   ~ &  ~ &  ~ \\
 5208.339 &  42.3  &  1.0  &  5208.427  &  46.4  &  1.3  &  5208.582  &  29.2  &  1.5  &   ~ &  ~ &  ~ \\
 5208.340 &  41.2  &  1.0  &  5208.428  &  49.2  &  1.3  &  5208.583  &  29.5  &  1.5  &   ~ &  ~ &  ~ \\
 5208.341 &  42.0  &  1.0  &  5208.429  &  49.4  &  1.3  &  5208.584  &  27.1  &  1.5  &   ~ &  ~ &  ~ \\
 5208.342 &  41.0  &  1.0  &  5208.430  &  50.8  &  1.3  &  5208.585  &  26.9  &  1.5  &   ~ &  ~ &  ~ \\
 5208.343 &  40.3  &  1.0  &  5208.431  &  48.5  &  1.3  &  5208.586  &  29.7  &  1.5  &   ~ &  ~ &  ~ \\
 5208.344 &  38.0  &  1.0  &  5208.432  &  46.3  &  1.3  &  5208.587  &  25.1  &  1.5  &   ~ &  ~ &  ~ \\

\hline
\end{tabular}
\end{table*} 

\clearpage
\newpage

\begin{table}[h!]
\caption{Julian day, flux, and uncertainty values for the millimeter light curve on 2010 January 11-12, recorded at a central frequency of 92.5\,GHz with CARMA. \label{jan10carma} } 
\centering 
\begin{tabular}{ccc|ccc}
\hline 
\hline
 ~Julian Day & Flux & $\sigma$ &  ~Julian Day & Flux & $\sigma$  \\
 ($+$2450000) & [mJy] & [mJy] & ($+$2450000) & [mJy] & [mJy]  \\
\hline
\noalign{\smallskip}

 5208.572  &  30.3  &	3.6  &  5208.843  &  18.6   &  3.4  \\
 5208.576  &  27.6  &	3.8  &  5208.847  &  18.1   &  4.0  \\
 5208.579  &  26.2  &	3.5  &  5208.850  &  25.3   &  4.0  \\
 5208.586  &  27.3  &	3.0  &  5208.857  &  17.4   &  4.1  \\
 5208.590  &  26.8  &	3.2  &  5208.860  &  14.9   &  4.4  \\
 5208.594  &  27.5  &	3.2  &  5208.864  &  18.1   &  4.9  \\
 5208.601  &  26.5  &	3.0  &  5208.879  &  19.9   &  4.9  \\
 5208.604  &  29.1  &	3.3  &  5208.883  &  23.3   &  6.2  \\
 5208.608  &  23.6  &	3.4  &  5208.886  &  21.5   &  5.3  \\
 5208.615  &  20.8  &	3.6  &  5208.893  &  18.5   &  5.2  \\
 5208.618  &  17.0  &	3.9  &  5208.897  &  13.1   &  5.5  \\
 5208.622  &  15.2  &	4.4  &  5208.900  &  11.2   &  5.7  \\
 5208.628  &  28.4  &	3.7  &  5208.907  &  9.4    &  5.9  \\
 5208.632  &  27.0  &	3.5  &  5208.910  &  9.6    &  6.2  \\
 5208.636  &  23.0  &	3.7  &  5208.914  &  10.1   &  6.1  \\
 5208.643  &  28.5  &	3.1  &  5208.921  &  12.4   &  7.7  \\
 5208.647  &  21.8  &	3.0  &   ~ &  ~ &  ~ \\
 5208.650  &  24.3  &	3.0  &   ~ &  ~ &  ~ \\
 5208.657  &  21.1  &	3.1  &   ~ &  ~ &  ~ \\
 5208.660  &  16.1  &	3.1  &   ~ &  ~ &  ~ \\
 5208.664  &  21.1  &	3.0  &   ~ &  ~ &  ~ \\
 5208.671  &  29.8  &	3.1  &   ~ &  ~ &  ~ \\
 5208.674  &  29.9  &	8.0  &   ~ &  ~ &  ~ \\
 5208.678  &  20.2  &	3.2  &   ~ &  ~ &  ~ \\
 5208.685  &  13.1  &	2.9  &   ~ &  ~ &  ~ \\
 5208.689  &  18.4  &	2.9  &   ~ &  ~ &  ~ \\
 5208.692  &  18.9  &	3.1  &   ~ &  ~ &  ~ \\
 5208.700  &  21.1  &	3.0  &   ~ &  ~ &  ~ \\
 5208.703  &  29.2  &	3.1  &   ~ &  ~ &  ~ \\
 5208.707  &  19.1  &	3.0  &   ~ &  ~ &  ~ \\
 5208.714  &  24.7  &	2.8  &   ~ &  ~ &  ~ \\
 5208.717  &  25.6  &	7.7  &   ~ &  ~ &  ~ \\
 5208.721  &  20.3  &	4.0  &   ~ &  ~ &  ~ \\
 5208.728  &  24.4  &	3.5  &   ~ &  ~ &  ~ \\
 5208.732  &  22.0  &	3.8  &   ~ &  ~ &  ~ \\
 5208.735  &  17.8  &	4.1  &   ~ &  ~ &  ~ \\
 5208.743  &  23.4  &	3.7  &   ~ &  ~ &  ~ \\
 5208.747  &  30.9  &	3.8  &   ~ &  ~ &  ~ \\
 5208.750  &  24.8  &	3.8  &   ~ &  ~ &  ~ \\
 5208.757  &  21.3  &	3.4  &   ~ &  ~ &  ~ \\
 5208.760  &  19.9  &	3.4  &   ~ &  ~ &  ~ \\
 5208.765  &  19.5  &	3.3  &   ~ &  ~ &  ~ \\
 5208.772  &  22.0  &	3.1  &   ~ &  ~ &  ~ \\
 5208.775  &  21.1  &	3.2  &   ~ &  ~ &  ~ \\
 5208.778  &  18.7  &	3.5  &   ~ &  ~ &  ~ \\
 5208.786  &  2.8   &	3.1  &   ~ &  ~ &  ~ \\
 5208.790  &  25.1  &	5.7  &   ~ &  ~ &  ~ \\
 5208.794  &  18.9  &	4.3  &   ~ &  ~ &  ~ \\
 5208.801  &  16.7  &	3.2  &   ~ &  ~ &  ~ \\
 5208.804  &  22.4  &	3.5  &   ~ &  ~ &  ~ \\
 5208.808  &  17.0  &	3.7  &   ~ &  ~ &  ~ \\
 5208.815  &  21.0  &	3.5  &   ~ &  ~ &  ~ \\
 5208.819  &  15.6  &	3.5  &   ~ &  ~ &  ~ \\
 5208.822  &  1.3   &	3.7  &   ~ &  ~ &  ~ \\
 5208.829  &  9.7   &	3.4  &   ~ &  ~ &  ~ \\
 5208.833  &  20.9  &	3.6  &   ~ &  ~ &  ~ \\
 5208.836  &  4.0   &	3.7  &   ~ &  ~ &  ~ \\
			      
\hline			      
\end{tabular}		      
\end{table}

\begin{table}[h!]
\caption{Julian day, flux, and uncertainty values for the millimeter light curve on 2010 January 12, recorded at a central frequency of 238.5\,GHz with the SMA. \label{jan10sma} } 
\centering 
\begin{tabular}{ccc}
\hline 
\hline
 ~Julian Day & Flux & $\sigma$  \\
 ($+$2450000) & [mJy] & [mJy]  \\
\hline
\noalign{\smallskip}

 5208.612  &  49.7   &  8.7   \\
 5208.619  &  28.2   &  7.3   \\
 5208.628  &  50.2   &  7.4   \\
 5208.640  &  62.0   &  6.8   \\
 5208.648  &  66.0   &  6.3   \\
 5208.656  &  92.7   &  6.6   \\
 5208.664  &  96.9   &  6.1   \\
 5208.676  &  98.5   &  5.9   \\
 5208.684  &  97.1   &  5.7   \\
 5208.692  &  91.4   &  5.7   \\
 5208.700  &  82.8   &  5.6   \\
 5208.712  &  94.8   &  5.5   \\
 5208.740  &  98.9   &  5.3   \\
 5208.747  &  103.0  &  5.7   \\
 5208.756  &  100.1  &  5.2   \\
 5208.767  &  100.0  &  5.2   \\
 5208.776  &  98.6   &  5.2   \\
 5208.784  &  93.5   &  5.1   \\
 5208.792  &  96.6   &  5.2   \\
 5208.803  &  89.4   &  5.3   \\
 5208.836  &  92.6   &  5.5   \\
 5208.844  &  110.6  &  5.1   \\
 5208.853  &  104.6  &  5.2   \\
 5208.865  &  106.2  &  5.2   \\
 5208.873  &  101.4  &  5.3   \\
 5208.881  &  106.8  &  5.4   \\
 5208.889  &  94.1   &  4.9   \\
 5208.901  &  102.4  &  5.0   \\
 5208.910  &  97.5   &  5.3   \\
 5208.917  &  94.0   &  5.2   \\
 5208.926  &  95.0   &  5.3   \\
 5208.938  &  96.0   &  5.4   \\
 5208.965  &  91.3   &  5.5   \\
 5208.973  &  99.7   &  5.8   \\
 5208.981  &  102.8  &  6.0   \\
 5208.993  &  97.5   &  6.0   \\
 5209.001  &  98.4   &  6.5   \\
 5209.009  &  89.6   &  6.4   \\
 5209.017  &  85.5   &  7.1   \\
 5209.027  &  80.5   &  7.6   \\
			      
\hline			      
\end{tabular}		      
\end{table}

\end{appendix}

\end{document}